\documentclass[pra,floatfix,twocolumn,showpacs]{revtex4}

\usepackage{amsmath,amssymb,amsfonts}
\usepackage{float}
\usepackage{graphicx} 
\usepackage{dcolumn} 
\usepackage{bm} 
\usepackage{bbm} 

\newcommand{\erf}[1]{Eq.~(\ref{#1})}
\newcommand{\erfs}[2]{Eqs.~(\ref{#1}) and (\ref{#2})}

\newcommand{\rt}[1]{\sqrt{#1}\,}

\newcommand{\an}[1]{\left\langle{#1}\right\rangle}
\newcommand{\ban}[1]{\big\langle{#1}\big\rangle}

\newcommand{\bTr}[1]{\Tr\big[{#1}\big]}

\newcommand{\tbt}[4]{\left(\begin{array}{cc} {#1} & {#2} \\[0.10cm] {#3} & {#4} \end{array}\right)}
\newcommand{\tbo}[2]{\left(\begin{array}{c} {#1} \\[0.05cm] {#2} \end{array}\right)}
\newcommand{\obt}[2]{\left(\begin{array}{cc} {#1} ,\; {#2} \end{array}\right)}

\newcommand{\vop}[1]{\hat{\bf {#1}}}
\newcommand{\vog}[1]{\hat{\bm{#1}}}

\newcommand{\smallfrac}[2]{\mbox{$\frac{#1}{#2}$}}

\newcommand{\ID}[1]{{\rm I}_{\text{\tiny $#1$}}}
\newcommand{\mopID}[1]{\hat{\rm I}_{\text{\tiny $#1$}}}
\newcommand{\mbrac}[2]{\big\lfloor {#1},{#2} \big\rceil}

\newcommand{\beq}{\begin{equation}} 
\newcommand{\eeq}{\end{equation}}
\newcommand{\bqa}{\begin{eqnarray}} 
\newcommand{\eqa}{\end{eqnarray}}
\newcommand{\nn}{\nonumber}
\newcommand{\half}{\smallfrac{1}{2}}

\newcommand{\ito}{It\^o} 

\newcommand{\sch}{Schr\"odinger} 
\newcommand{\hei}{Heisenberg}

\newcommand{\tp}{^{\top}} 
\newcommand{\dg}{^\dagger}
\newcommand{\ddg}{^\ddag}
\newcommand{\Tr}{{\rm Tr}}
\newcommand{\tr}{{\rm tr}}

\newcommand{\hc}{\hat{\bf c}} 
\newcommand{\vhc}{\hc_{\&}}
\newcommand{\rhoc}{\rho_{\rm c}}

\newcommand{\bin}{\vop{b}_{\rm in}}
\newcommand{\bout}{\vop{b}_{\rm out}}

\newcommand{\dBin}{d\vop{B}_{\rm in}}
\newcommand{\dBout}{d\vop{B}_{\rm out}}

\newcommand{\dups}{d\vog{\upsilon}}

\newcommand{\mrep}{{\sf M}}
\newcommand{\brep}{{\sf B}}
\newcommand{\urep}{{\sf U}}
\newcommand{\trep}{{\sf T}}
\newcommand{\MdgM}{{\sf M}\dg {\sf M}}
\newcommand{\MMdg}{{\sf M} {\sf M}\dg}

\begin{document}

\title{Complete Parameterizations of Diffusive Quantum Monitorings}

\author{A. Chia}
\author{H. M. Wiseman}
\affiliation{Centre for Quantum Computation and Communication Technology (Australian Research Council); \\
Centre for Quantum Dynamics, Griffith University, Brisbane, Queensland 4111, Australia}

\date{\today}

\begin{abstract}

The master equation for the state of an open quantum system can be unravelled into stochastic trajectories described by a stochastic master equation. Such stochastic differential equations can be interpreted as an update formula for the system state conditioned on results obtained from monitoring the bath. So far only one parameterization (mathematical representation) for arbitrary diffusive unravellings (quantum trajectories arising from monitorings with Gaussian white noise) of a system described by a master equation with $L$ Lindblad terms has been found [H. M. Wiseman and A. C. Doherty, Phys. Rev. Lett. {\bf 94}, 070405 (2005)]. This parameterization, which we call the U-rep, parameterizes diffusive unravellings by $L^2+2L$ real numbers, arranged in a matrix \urep\ subject to three constraints. In this paper we investigate alternative parameterizations of diffusive measurements. We find, rather surprisingly, the description of diffusive unravellings can be unified by a single equation for a non-square complex matrix if one is willing to allow for some redundancy by lifting the number of real parameters necessary from $L^2+2L$ to $3L^2+L$. We call this parameterization the M-rep. Both the M-rep and U-rep lack a physical picture of what the measurement should look like. We thus propose another parameterization, the B-rep, that details how the measurement is implemented in terms of beam-splitters, phase shifters, and homodyne detectors. Relations between the different representations are derived.

\end{abstract}

\maketitle

\section{Introduction}

Methods of detection for optical fields such as the heterodyne (and homodyne) \cite{SZ97} enable measurements of the field amplitude (and quadratures) and are useful for detecting nonclassical light \cite{CHG+00,CN04}. The heterodyne (or homodyne) detector is an example of general diffusive measurements \cite{WM10,BG09}, so-called because the statistical fluctuations in the current outputed by the measurement is driven by a Gaussian-white-noise process, a prime example of a diffusion process. This is the process responsible for the broadening of a probability distribution described by the diffusive term appearing in Fokker-Planck equations \cite{Gil92,Jac10}. The heterodyne technique, originally conceived for radio technology more than 100 years ago \cite{Bri04,Bro08}, of which the homodyne is a variant, has now become an indispensable part of many quantum information processing applications. Some of these applications include quantum-state tomography \cite{LR09,SBR+93,OTG06}, teleportation \cite{Vai94,BK98,FSB+98,ZGC+03}, and state preparation \cite{RK03,JKR+04,OJR+07,RJG+08,OJ09,LKH+10}. It also enables tests of quantum mechanics \cite{KWM00,NC04,GFC+04}.

In this paper we are concerned with continuous measurements (also referred to as monitorings) of an open system in a vacuum bosonic environment (Fig.~\ref{OverallSchemeSchPic}). Specifically we are interested in how one can specify the class of all possible diffusive quantum measurements given how many inputs to the measuring device there are. There is already one solution to this problem \cite{WD01,WD05}, a parameterization that we shall call the U-rep. The U-rep parameterizes an arbitrary diffusive measurement by a square, real matrix \urep, with three constraints imposed on the blocks of \urep. We will present two more solutions with certain advantages over the U-rep: The M-rep --- a non-square complex matrix \mrep; and the B-rep --- a square matrix and two vectors. We will be comparing the different parameterizations in detail later but let us first say why one might consider these alternative parameterizations.

A good reason, and also the more theoretical reason for considering the M-rep, is that aesthetically it is more attractive than the U-rep. As we will show, the M-rep has the ability to unify all diffusive measurements by a single equation whereas three would be required to define \urep\ \footnote{However, the M-rep is numerically more cumbersome than the U-rep in the sense that it requires more real numbers to specify a diffusive measurement. In particular, the extra number of parameters that \mrep\ needs scales as a quadratic in the number of rows of \mrep.}. The aesthetic advantange of the M-rep continues when we come to the quantum theory of multiple-input multiple-output Markovian feedback control based on diffusive measurements \cite{CW11b}. There, the theory is simpler when the equations are expressed in terms of the M-rep rather than the U-rep. Furthermore, the defining equation for \mrep\ has an intuitive interpretation which makes it easy to remember.


Either the U-rep or the M-rep can be used when we know what the measurement is and would like to parameterize it in order to model the state of the open system conditioned on the results of the monitoring. If on the other hand we were given either a \urep, or an \mrep, and were asked to describe the measurement with actual optical elements then this is a much more difficult task for arbitrary measurements. Such a setting where one may be given an \mrep\ without knowing its physical implementation can in fact arise naturally in quantum feedback control \cite{WD05,MW07}: As measurement is an inherent part of the feedback loop, a design of the feedback loop can thus incorporate a design of measurement and it is natural to ask what sort of measurement one should do in order to achieve a particular objective for the control. In the case of optimal control the objective would be to minimize a measurement-dependent cost function. The result of this optimization would be a matrix $\mrep^\star$, which one would then need to realize in the laboratory. That is, the theorist who now has obtained $\mrep^\star$ would like to inform his/her experimentalist colleague about how to construct the measurement.

The above considerations motivate us to propose yet another parameterization of diffusive measurements; one which we call the block-rep, or B-rep for short. Unlike the M- and U-rep, the B-rep is a realization of the measurement in terms of beam-splitters, phase shifters and homodyne detectors. The B-rep is so-called because it parameterizes the diffusive measurement in terms of three distinct blocks, with each block corresponding to a set of parameters (beam-splitter transmission coefficients and phase shifts). Note that being given an \mrep\ is equivalent to being given a POVM (positive-operator-valued-measure \footnote{The outcomes of a measurement can be labelled by a random variable ${\bf y}$ (here assumed to be discrete). In a general formulation of quantum measurements the distribution of the particular outcomes $\breve{\bf y}$ (a deterministic variable) of ${\bf y}$ is described by a set of ``probability operators'' \cite{WM10,NC10}, defined by $\big\{\hat{E}_{\breve{\bf y}} \,|\, \forall \,\breve{\bf y}\; \hat{E}_{\breve{\bf y}} \ge 0\,,\; \sum_{\breve{\bf y}} \hat{E}_{\breve{\bf y}} = \hat{1} \big\}$. This set is known as a POVM and sometimes also as a POM, short for probability-operator-measure. We shall find in Sec.~\ref{DerivanOfMeasRecord} that ${\bf y}$ depends on \mrep\ (see \eqref{MrepCurrent}) so that a given \mrep\ will restrict the set of possible realizations $\breve{\bf y}$. Each choice of \mrep\ therefore defines a POVM and each value of $\breve{\bf y}$ corresponds to a particular POVM element $\hat{E}_{\breve{\bf y}}$.}). From this point of view the relationship between the M-rep and B-rep is thus one of a POVM and its realization. As we shall see, proving that an arbitrary POVM has a realization defined by the B-rep is much more difficult than translating a given B-rep into an \mrep, and hence a POVM. Here we point out that this line of thought had in fact been applied to $N$-port homodyne detection --- a generalization of the standard homodyne measurement but a subset of all diffusive measurements, as early as 1987 \cite{Wal87}. A similar question was raised again in 1994, but for unitary operators instead of a POVM \cite{RZB+94}. Our construction of the B-rep in fact relies on this result of Ref.~\cite{RZB+94}.

Our paper is organized as follows. In Sec.~\ref{BackgroundToQuantumTrajectories} the unconditional dynamics of open quantum systems is reviewed and the concept of unravellings and its relation to continuous measurements are also briefly sketched. As we frequently make use of vector-operators, henceforth referred to as vops, a comprehensive review of their definitions and algebra is provided in Appendix~\ref{VecOperatorAlg}. Note that not all of the results in Appendix~\ref{VecOperatorAlg} will be used in this paper so we will refer to the ones that do appear. However, the reader who is also interested in the theory of multiple-input multiple-output quantum feedback control (for which the M-rep is applied to) \cite{CW11b} may find it worthwhile to embark on a fuller exposition of vop algebra. We also mention that for convenience we will not necessarily reflect the multi-component nature of vectors or vops in our language when they are referred to, such as in ``the field $\vop{a}$'', or, ``the current $\vop{y}$'', as opposed to using plurals as in ``the fields $\vop{a}$'' or ``the currents $\vop{y}$''. Our first results section begins with Sec.~\ref{M-Representation}. The key results here are the definition of the M-rep and its relation with the U-rep. It should be noted that Sec.~\ref{M-Representation} begins by postulating the most general diffusive stochastic master equation. The purpose of Sec.~\ref{BackgroundToQuantumTrajectories} is thus to help the reader gain the intuition required to make this leap. In Sec.~\ref{BlockRepresentation} we define the B-rep and discuss its relation with the M-rep. Here we conjecture a matrix decomposition of \mrep\ in terms of the matrices in the B-rep triple. We then conclude in Sec.~\ref{ParameterizationConclusion}.


\section{Diffusive Continuous Measurements in the \sch\ Picture}
\label{BackgroundToQuantumTrajectories}

\subsection{Unconditional System Dynamics}
\label{SchPicParamUnrav}

\begin{figure}
\includegraphics[width=8.4cm]{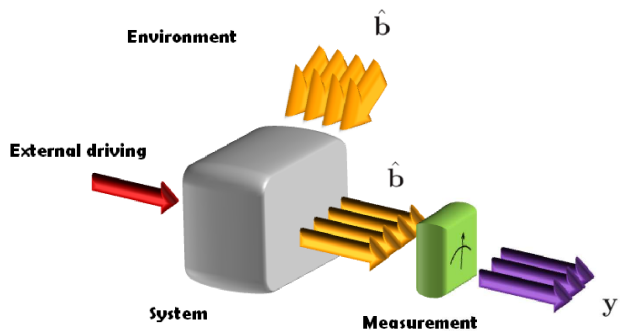} 
\caption{Schematic of some photoemissive system which couples to the environment via some set of operators $\hc$. The time evolution of the system based on knowledge of the noisy signal $y$ also follows a noisy path, which can be described by a stochastic master equation. The stochastic master equation updates the system state continuously in time as the measurement record grows. If one ignores the measured result $y$ then the system state evolves according to the deterministic master equation. This corresponds to averaging over the stochastic evolution.}
\label{OverallSchemeSchPic}
\end{figure}

A general open quantum system with Markovian dynamics \cite{AL07,BP02} can be described by a master equation in the Lindblad form \cite{Lin76,Car02}
\begin{align}
\label{LindbladForm}
	\hbar \, \dot{\rho} = -i \;\! [\hat{H},\rho] + {\cal D}[\hat{\bf c}] \rho \equiv {\cal L} \;\! \rho 
\end{align}
where $\hat{H}$ is Hermitian and $\hat{\bf c} \equiv (\hat{c}_{1},\hat{c}_{2},\ldots,\hat{c}_{L})\tp$ (which we will refer to as a vop) is a vector of Lindblad operators, all time-independent. See Appendix~\ref{VecOperatorAlg} for a comprehensive review of definitions and conventions that we are adopting for vops. We have also defined 
\begin{align}
\label{SuperD}
	{\cal D}[\hat{\bf c}] \equiv \sum^{L}_{k=1} {\cal D}[\hat{c}_{k}] \,, \;  
	{\cal D}[\hat{c}] \rho \equiv \hat{c} \;\! \rho \;\! \hat{c}\dg
	                              - \half \;\! \hat{c}\dg \hat{c} \rho - \half \;\! \rho \;\! \hat{c}\dg \hat{c} \,.
\end{align}
Note that instead of working in natural units by setting $\hbar=1$ we will work in units such that a factor of $\hbar$ appears on the LHS of \eqref{LindbladForm}. This was done in Refs.~\cite{WD05,WM10} so we will keep their convention for ease of comparison with the results therein. The reason for redefining units this way is to keep track of the correspondence between quantum operators and their classical counterpart (the dynamical variables which they represent) and aspects of quantum control that are not present in the classical theory. Equation \eqref{LindbladForm} can be derived by a unitary operator $\hat{U}(t,t_0)$ acting on the joint Hilbert space of the system and bath \cite{BP02}. This describes the interaction between the system and its environment, which we assume to be given by the \ito\ stochastic differential equation 
\begin{align}
\label{HudsonParthasarathySch}
	\hbar \, d\hat{U}(t,t_0) =  \big( \!-\!i\hat{H} - \smallfrac{1}{2} \;\! \hc\dg \hc  
	                            + \hat{\bf b}\dg \hc - \hc\dg \hat{\bf b} \;\!\big) \hat{U}(t,t_0) \, dt \;,
\end{align}
where $d\hat{U}(t,t_0) \equiv \hat{U}(t+dt,t_0)-\hat{U}(t,t_0)$ and 
\begin{align}
\label{VopConjuageDefn}
	\hc\dg \equiv \big(\hat{c}\dg_1,\hat{c}\dg_2,\ldots,\hat{c}\dg_L\big) \;.
\end{align}
Here $\hat{\bf b}$ represents a bosonic bath field shown in Fig.~\ref{OverallSchemeSchPic} and assumed to be in the vacuum state. External driving of the system is included in $\hat{H}$. In this section we are working in the \sch\ picture so operators are time-independent, equal to their initial value.

It is well known that the evolution described by \eqref{LindbladForm} can be decomposed into stochastic paths in Hilbert space, called quantum trajectories (a term coined by Carmichael \cite{Car93} who also referred to this decomposition as unravelling the master equation) by considering the evolution of $\rho$ conditioned on the results of monitoring \cite{Car93,Car08}. We will also refer to the results of monitoring as a current, denoted by a vector ${\bf y}(t)$. For Markovian evolution it is sufficient to consider unravellings (the set of possible quantum trajectories the system state may take over time) generated by measurements with either or both of two classes of noise \cite{Gar85,WM10}: 1) A point (or jump) process \cite{PK98}, also called a Poissonian random variable, or; 2) a Wiener process also called Gaussian white noise. We will refer to this second case as a diffusive measurement. Physicists usually neglect the technical difference between an ``unravelling'' and a ``measurement,'' and one often finds the two terms used interchangeably. We will do so as well in this paper for convenience. 
%
%

\subsection{Conditional System Dynamics --- Homodyne and Heterodyne Unravellings}
\label{NotationForQuantumTrajectories}

The most familiar forms of diffusive measurements in quantum optics are homodyne and heterodyne detection schemes of a single output field of some photoemissive source. Apart from being characterized by Gaussian white noise (as opposed to point-process noise) these measurements are also determined by physical properties such as the specific arrangement of linear optical elements and the specific parameter values at which they operate. Here we will first show how the master equation \eqref{LindbladForm} is unravelled by the homodyne and heterodyne detection schemes. This will familiarize the reader with the notation used for describing conditional evolution and build some intuition about how the measurement parameters enter into a quantum trajectory equation. For convenience we will set $\hbar=1$ for the rest of this section.

\subsubsection{The homodyne stochastic master equation}

We begin with the example of a single mode of an intracavity field. For simplicity we will assume the optical cavity to be one-ended with transmission coefficient $\gamma$ and that the bath is in the vacuum state. In this case the master equation \eqref{LindbladForm} has only one Lindblad term (i.e. $L=1$), given by $\hat{c} = \rt{\gamma} \hat{a}$, where $\hat{a}$ is the cavity mode annihilation operator. A homodyne measurement \cite{SZ97} of the field at the leaky port would then give us information about a particular quadrature of $\hat{a}$. Such a continuous measurement would yield a result, for some state $\rho(t)$, given by the \ito\ formula \cite{WM10,Wis96} (assuming a measurement efficiency of $\eta$)
\begin{align}
\label{SimpleHomodyneMmt}
	y(t) \;\! dt = \rt{\eta} \an{\hat{c} + \hat{c}\dg}\!(t) \, dt + dw(t) \;.
\end{align}
We have defined, for an arbitrary operator $\hat{A}$, $\langle\hat{A}\rangle(t)={\rm Tr}[\hat{A}\;\!\rho(t)]$, reserving angle brackets for quantum (operator) averages. The term $dw(t)$ is a Wiener increment. It is a Gaussian random variable defined by the mean
\begin{align}
\label{ClassicalAvgNotation}
	{\rm E}[ dw(t) ] = 0
\end{align}
and variance
\begin{align}
	[ dw(t) ]^2 = {}& dt \;, \\
	dw(t) \;\! dw(t') = {}& 0  \quad  \forall \; t \ne t' \;.
\end{align}
We shall always denote an average taken with respect to a probability distribution by an E as in \eqref{ClassicalAvgNotation}.

From the measurement result $y(t)$ the observer can update her \emph{prior} state from $\rho(t)$ to a \emph{posterior} state $\rho_{y_t}(t+dt)$ according to the stochastic master equation
\begin{align}
\label{SimpleNonlinearSME}
	d\rho_{y_t}(t) = {\cal L} \;\! \rho(t) \, dt + \rt{\eta} \, {\cal H}[\hat{c}] \;\! \rho(t) \, dw  \;,
\end{align}
where ${\cal L}$ is given by \eqref{LindbladForm} and the change in the state is
\begin{align}
\label{SingleTimeConditioning}
	d\rho_{y_t}(t) = \rho_{y_t}(t+dt) - \rho(t) \;.
\end{align}
We have also defined (see also \eqref{SuperOpH1} in Appendix~\ref{VecOperatorAlg})
\begin{align}
\label{SuperOpH1MainText}
	{\cal H}[\hat{A}] \hat{B} \equiv \hat{A} \hat{B} + \hat{B} \hat{A}\dg - \Tr \big( \hat{A} \hat{B} + \hat{B} \hat{A}\dg \big) \hat{B} \;,
\end{align}
for any $\hat{A}$ and $\hat{B}$. Here $\rho_{y_t}(t+dt)$ is the state given a particular realization of $y(t)$. We have also written the time-dependence of $y$ as a subscript. For clarity it is best to have this flexibility in writing the time-dependence. Averaging \eqref{SimpleNonlinearSME} over $y(t)$ according to its actual distribution $\wp(\breve{y}_t)$ (a Gaussian with mean $\rt{\eta} \an{\hat{c} + \hat{c}\dg}$ and variance $1/dt$) simply returns the master equation \eqref{LindbladForm}. Note that a breve over a random variable denotes its realization. 

Equation \eqref{SimpleNonlinearSME} is a nonlinear stochastic differential equation in the state. This would not have been the case if we did not insist on keeping the state normalized at all times and having the measurement results distributed according to its true statistics described by $\wp(\breve{y}_t)$. An alternative theory is to assign an ostensible distribution $\wp_{\rm ost}(\breve{y}_t)$ to $y(t)$ and just disregard the norm of the state. When $\wp_{\rm ost}(\breve{y}_t)$ is a Gaussian with zero mean and variance $1/dt$ the state evolves according to a \emph{linear} stochastic master equation given by
\begin{align}
\label{SimpleNonLinearSME}
	d\bar{\rho}_{y_t}(t) = {\cal L} \;\! \rho(t) \, dt +  y(t) \;\! dt \, \bar{\cal H}[\hat{c}] \;\! \rho(t) \;,
\end{align}
where the updated state is now unnormalized
\begin{align}
\label{SingleTimeConditioningUnnorm}
	d\bar{\rho}_{y_t}(t) = \bar{\rho}_{y_t}(t+dt) - \rho(t) \;,
\end{align}
and we have defined a ``linear'' version of \eqref{SuperOpH1MainText}
\begin{align}
	\bar{\cal H}[\hat{A}] \hat{B} \equiv {}& \hat{A} \hat{B} + \hat{B} \hat{A}\dg  \;.
\end{align}
The solutions to \eqref{SimpleNonLinearSME} are called linear quantum trajectories \cite{Wis96}. For linear quantum trajectories to be equivalent to the standard theory (the case when the measurement results are distributed according to their true distribution $\wp(\breve{y}_t)$ and the state follows a nonlinear stochastic master equation) the ostensible distribution must be such that \cite{WM10}
\begin{align}
\label{TrueOstensibleRelation}
	\wp(\breve{y}_t) \, d\breve{y}_t = {\rm Tr}\big[\bar{\rho}_{y_t}(t+dt)\big] \: \wp_{\rm ost}(\breve{y}_t) \, d\breve{y}_t \;.
\end{align}
The form of the linear stochastic master equation is specific to the form of $\wp_{\rm ost}(\breve{y}_t)$, so choosing a different form for $\wp_{\rm ost}(\breve{y}_t)$ will result in a different linear stochastic master equation \cite{WM10}.

In the above we have assumed the prior state to be unconditioned (implied by \eqref{SingleTimeConditioning}), no matter how it is obtained. In general this need not be the case, and certainly will not be once the prior state is propagated for any finite time interval under \eqref{SimpleNonLinearSME} (or \eqref{SimpleNonlinearSME}). Thus for generality and clarity we will introduce a Roman subscript c (for ``conditioned'') and write the linear stochastic master equation as    
\begin{align}
\label{SloppyNotationSME}
	d\bar{\rho}_{\rm c}(t) = {\cal L} \, \rho(t) \, dt +  y(t) \;\! dt \: \bar{\cal H}[\hat{c}] \;\! \rho(t) \;,
\end{align}
where 
\begin{align}
\label{SloppyNotationStateChange}
	d\bar{\rho}_{\rm c}(t) = \bar{\rho}_{\rm c}(t+dt) - \rho(t) \;.
\end{align}
Tacked onto \erfs{SloppyNotationSME}{SloppyNotationStateChange} is the system of notation that $\rho(t)$ denotes a state of knowledge that is arbitrarily conditioned on the measurement record during $[0,t)$. Once $\rho(t)$ is determined the precise conditioning of $\bar{\rho}_{\rm c}(t+dt)$ can then be expressed. We will inform the reader about the dependence of $\rho(t)$ on the measurement record when it is necessary to assume a particular conditioning for $\rho(t)$.

\subsubsection{The heterodyne stochastic master equation}
\label{Example-HeterodyneSME}

A heterodyne measurement of efficiency $\eta$ can be shown to be formally equivalent to two simultaneous measurements of orthogonal quadratures of the signal field by two homodyne detectors each with efficiency $\eta/2$. This latter detection scheme of using two homodyne detectors is also referred to as a dual-homodyne measurement. It is a convenient way of understanding the heterodyne measurement and is how we will think about it. That is, while we use the term ``heterodyne'' often, we will always realize it using a dual-homodyne scheme. This decomposition of a heterodyne measurement into two homodyne ones will also be of use later in Sec.~\ref{BlockRepresentation}.

If we now keep the above example of a leaky cavity and perform a heterodyne (or dual-homodyne) measurement of the output of the cavity then we obtain two currents, $y_1$ and $y_2$, which we write as the components of a $2 \times 1$ vector ${\bf y}$,
\begin{align}
\label{HeterodyneCurrent}
	{\bf y} \, dt = \rt{\frac{\eta}{2}} \tbo{\ban{\hat{c}+\hat{c}\dg}}{-i\ban{\hat{c}-\hat{c}\dg}} dt + d{\bf w}  \;.
\end{align}
where $d{\bf w}=(dw_1,dw_2)\tp$ and $dw_1$ and $dw_2$ are independent Wiener increments,
\begin{align}
	dw_1(t) \;\! dw_2(t') = 0  \quad \forall \; t,t' \;.
\end{align}
Conditioning the system state on \eqref{HeterodyneCurrent} then leads to the heterodyne stochastic master equation
\begin{align}
\label{HeterodyneSME}
	d\rhoc(t) = {}& {\cal L} \, \rho(t) \, dt  \nn \\
	              & + \rt{\frac{\eta}{2}} \, {\cal H}[\hat{c}] \;\! \rho(t) \, dw_1 
	                + \rt{\frac{\eta}{2}} \, {\cal H}[-i\hat{c}\;\!] \;\! \rho(t) \, dw_2  \;.
\end{align}
We can write this more compactly by defining the row vector
\begin{align}
	{\sf M} = \rt{\frac{\eta}{2}} \, (1,i) \;.
\end{align}
Equation \eqref{HeterodyneSME} is then equivalent to
\begin{align}
	d\rhoc(t) = {\cal L} \, \rho(t) \, dt + d{\bf w}\tp {\cal H}[\mrep\dg \hat{c}\;\!] \;\! \rho(t)  \;.
\end{align}
Note here the use of the vop-valued superoperator \eqref{SuperOpH2}, defined for any $\vop{A}$ and $\hat{B}$ as
\begin{align}
\label{VopvaluedCalH}
	{\cal H}[\vop{A}] \hat{B} \equiv \vop{A} \hat{B} + \hat{B} \vop{A}\ddg - \Tr \big[ \vop{A} \hat{B} + \hat{B} \vop{A}\ddg \big] \hat{B}  \;,
\end{align}
where $\vop{A}\ddg \equiv \big( \vop{A}\tp \big)\dg$. The heterodyne measurement illustrates how ${\bf y}$ can be a two-component vector even though $L=1$ so there is only one output field $\hat{b}_{\rm out}$.

\section{The M-representation}
\label{M-Representation}

In this section we show how a single complex matrix can capture any such properties of diffusive measurements of an arbitrary number of system outputs in the \sch\ picture. In Sec.~\ref{BlockRepresentation} a second and physically intuitive method of parameterizing diffusive unravellings is formulated in the \hei\ picture.

\subsection{Measurement Statistics}
\label{DerivanOfMeasRecord}

Based on the foregoing examples of homodyne and heterodyne unravellings in Sec.~\ref{NotationForQuantumTrajectories} we propose the following form of a general diffusive stochastic master equation 
\begin{align}
\label{GeneralNonlinearDiffSME}
	\hbar \, d\rho_{\rm c}(t) = {\cal L} \;\! \rho(t) \, dt + d{\bf w}\tp {\cal H}[ {\sf M}\dg \hc \;\!] \;\! \rho(t) \;,
\end{align}
conditioned on the $2L \times 1$ real vector
\begin{align}
\label{GeneralDiffMmtRecord}
	{\bf y}(t) \;\! dt =  {\rm E}\big[ {\bf y}(t) \big] dt + d{\bf w}(t) \;,
\end{align}
which represents the measured current. The length of ${\bf y}$ is motivated by the example of heterodyne detection in Sec.~\ref{Example-HeterodyneSME} and the fact that here we are allowing $L$ components in $\bout$. The measurement is a noisy process. The noise in the measurement is given by a vector of independent Wiener increments $d{\bf w}$ (also $2L \times 1$ by default). The vector $d{\bf w}(t)$ therefore has zero mean 
\begin{align}
\label{ClassicalIto1}
	{\rm E}\big[ d{\bf w}(t) \big] = {\bf 0} \quad \forall \, t \;,
\end{align}
and correlation matrices
\begin{align}
\label{ClassicalIto2}
	d{\bf w}(t) \, d{\bf w}\tp(t) = {}& \ID{2L} \, dt  \;, \\
\label{ClassicalIto3}
	d{\bf w}(t) \, d{\bf w}\tp(t') = {}& 0  \quad  \forall \;  t \ne t'  \;.
\end{align} 
In \eqref{GeneralNonlinearDiffSME} we have used \eqref{VopvaluedCalH} and defined ${\sf M}$ to be an $L \times 2L$ complex matrix that parameterizes the unravelling \footnote{In the case when some of the columns of \mrep\ are zero, say $R$ ($\le 2L$), then $d{\bf w}$ can be defined to be $R \times 1$ instead of $2L \times 1$ (in fact we do this in Ref.~\cite{CW11b}). Defining \mrep\ to be $L \times 2L$ and $d{\bf w}$ to be $2L \times 1$ here just simplifies our theory.}. The constraints which define ${\sf M}$ will be determined in the following sections, here we will obtain the form of the measured current consistent with \eqref{GeneralNonlinearDiffSME}.

The precise of form of ${\rm E}\big[{\bf y}(t)\big]$ can in fact be derived from the theory of linear quantum trajectories which allows us to calculate the actual statistics of the measured current by using an ostensible distribution in the following way: Here we are only concerned with the average of ${\bf y}_t$ and this is 
\begin{align}
\label{AvgFromLinearTraj}
	{\rm E}\big[ {\bf y}(t) \big] dt = \bar{\rm E}\big[ \;\! {\bf y}(t) \;\! {\rm Tr}\big\{ \rho(t) + d\bar{\rho}_{\rm c}(t) \big\} \big] \;\! dt \;,
\end{align}
where $\rho(t)$ is normalized and $d\bar{\rho}_{\rm c}(t)$ is given by
\begin{align}
\label{GeneralLinearDiffSME}
	\hbar \, d\bar{\rho}_{\rm c}(t) = {\cal L} \;\! \rho(t) \, dt + {\bf y}\tp(t) \bar{\cal H}[{\sf M}\dg \hc ] \;\! \rho(t) \;\! dt  \,.
\end{align}
Here we have used \eqref{TrueOstensibleRelation} to rewrite the average on the LHS as an average with respect to the ostensible distribution (denoted by an overbar on E)
\begin{align}
\label{MultiDimensionalOstDist}
	\wp_{\rm ost}(\breve{\bf y}_{t}) 
	= \left(\frac{dt}{\rt{2\pi}}\right)^{L} \; \exp\!\big[ \!-\! \half \, \breve{\bf y}\tp_{t} \breve{\bf y}_{t} \, dt \big] \;.
\end{align}
This is a Gaussian with mean zero and covariance
\begin{align}
	\bar{\rm E}\big[ {\bf y}(t) \, {\bf y}\tp\!(t) \big] = \ID{2L} / dt \;.
\end{align}
Defining a linear version of \eqref{VopvaluedCalH},
\begin{align}
	{\cal H}[\vop{A}] \hat{B} \equiv \vop{A} \hat{B} + \hat{B} \vop{A}\ddg  \;,
\end{align}
we then obtain, on substituting \eqref{GeneralLinearDiffSME} into \eqref{AvgFromLinearTraj},
\begin{align}
\label{TrueyAvg1}
	{\rm E}\big[ {\bf y}(t) \big] dt 
	= {}& \hbar^{-1} \, \bar{\rm E}\big[ \;\! {\bf y}(t) \;\! dt \; 
	      {\bf y}\tp\!(t) \;\! dt \, {\rm Tr}\big\{ \bar{\cal H}[{\sf M}\dg \;\! \hc] \;\! \rho(t) \big\} \big] \\
	= {}& \hbar^{-1} \, {\rm Tr}\big\{ {\sf M}\dg \hc \;\! \rho(t) + \rho(t) \;\! {\sf M}\tp \hc\ddg \big\}  \nn \\
\label{TrueyAvg2}
	= {}& \frac{dt}{\hbar} \; \big\langle {\sf M}\dg \hc + {\sf M}\tp \hc\ddg \big\rangle \;.
\end{align}
We have used the facts that ${\cal L}\rho$ is traceless and that $\bar{\rm E}[{\bf y}]={\bf 0}$ in \eqref{TrueyAvg1}. Later (Sec.~\ref{BlockRepresentation}) we will be considering a different representation of diffusive measurements. In order to distinguish between different representations we will use a subscript on the current. Substituting \eqref{TrueyAvg2} back into \eqref{GeneralDiffMmtRecord} we will thus write
\begin{align}
\label{MrepCurrent}
	\hbar \, {\bf y}_{\sf M} \, dt = \ban{{\sf M}\dg \hc + {\sf M}\tp \hc\ddg} dt + \hbar \, d{\bf w} \;.
\end{align}

It is clear from \eqref{TrueyAvg2} that ${\sf M}$ determines what property of the system gets measured, represented by some Hermitian operator $\hat{f}(\hc)$, and also the measurment statistics (i.e. the statistics of ${\bf y}_t$) for a measurement of $\hat{f}(\hc)$. That ${\sf M}$ defines the diffusive measurement could have been anticipated from \eqref{GeneralNonlinearDiffSME} since ${\sf M}$ appears in the ${\cal H}$ superoperator which stems from considering measurement. Its appearance in \eqref{GeneralNonlinearDiffSME} also means that not every ${\sf M}$ will be a valid parameterization of a diffusive measurement. In Sec.~\ref{DefiningM} we find the necessary and sufficient condition for ${\sf M}$ to be valid.

The careful reader who is familiar with the quantum theory of indirect measurements \cite{BK92,WM10} will know that the conditioning for the state is in general just a label which allows us to distinguish between measurement outcomes, and as such it does not have to be real. Our choice of conditioning the state on a real vector is motivated by its use in control theory in which the prevalent treatments regard ${\bf y}$ as a real process \cite{Lev96}.

The assumption that $\rho(t)$ is unnormalized can be relaxed. Say the measurement began at time $0$ and that $\bar{\rho}_{\rm c}(t)$ is obtained by evolving $\rho(0)$ via \eqref{GeneralLinearDiffSME} to time $t$. In this case the correct modification to the above calculation is to divide the trace in \eqref{AvgFromLinearTraj} by ${\rm Tr}\{ \bar{\rho}_{\rm c}(t) \}$. Keeping the same ostensible distribution as \eqref{MultiDimensionalOstDist} we get
\begin{align}
\label{ConditionedAvgFromLinearTraj}
	{\rm E}\big[ {\bf y}_t \big| \;\! {\bf y}_{[0,t)} \big] dt = \bar{\rm E}\!\left[ {\bf y}_t  
	                                                             \frac{ {\rm Tr}\big\{ d\bar{\rho}_{\rm c}(t) \big\} }
	                                                                  { {\rm Tr}\big\{ \bar{\rho}_{\rm c}(t) \big\} } \right]  dt \;,
\end{align}
where on the RHS the state $\bar{\rho}_{\rm c}$ is conditioned on the measurement record
\begin{align}
\label{DefnOfMmtRecord}
	{\bf y}_{[0,t)} \equiv \big\{ {\bf y}(s) \; | \; 0 \le s < t \big\}  \;.
\end{align}
The end result from calculating \eqref{ConditionedAvgFromLinearTraj} is to replace the unconditioned averages in \eqref{TrueyAvg2} by conditional ones,
\begin{align}
\label{ConditionedTrueAvgOfy}
	\langle \hc \rangle \longrightarrow \langle \hc \rangle_{\rm c} = \bTr{\hc \bar{\rho}_{\rm c}(t)} \;,                                      
\end{align}
as one would have guessed, where 
\begin{align}
	 \rho_{\rm c}(t) = \frac{\bar{\rho}_{\rm c}(t)}{{\rm Tr}[\bar{\rho}_{\rm c}(t)]} \;.
\end{align}

\subsection{The Set of Allowed \mrep}
\label{DefiningM}

\subsubsection{Necessary condition for ${\sf M}$}
\label{NecessConditionForM}

If ${\sf M}$ is to be a valid parameterization of a diffusive quantum measurement it must be such that \eqref{GeneralNonlinearDiffSME} evolves a valid state at time $t$ to another valid state $\rho(t+dt)$ for all $t$. Here we remind the reader that a valid state is represented by an Hermitian operator that is normalized and positive. From this it follows that its eigenvalues all lie in the interval $[0,1]$, which we can write as an operator-inequality
$	0 \le \rho \le \hat{1}$. The positivity condition is the only nontrivial criterion because ${\cal L}\rho$ and ${\cal H}\rho$ are each Hermitian and traceless so \eqref{GeneralNonlinearDiffSME} will always preserve Hermiticity and normalization for any ${\sf M}$. Since these conditions must hold for any measurement process, they are necessary for a given ${\sf M}$ to be valid, and we establish the following implication
\begin{align}
\label{ValidMStatement}
	\text{${\sf M}$ is valid}  \; \implies \;  {\sf M}: \, 0 \le \rho(t) + d\rho_{\rm c}(t) \le \hat{1} \;.
\end{align} 
where $\rho(t)$ is understood to be any valid state and $d\rho_{\rm c}(t)$ is given by \eqref{GeneralNonlinearDiffSME}. It is simple (given  $d\rho_{\rm c}(t)$ is traceless) to see that 
\begin{align}
\label{NecessaryImplication}
	0 \le \rho(t) + d\rho_{\rm c}(t) \le \hat{1}  \; \Longrightarrow \;  {\rm Tr}\big\{ [\rho(t) + d\rho_{\rm c}(t)]^2 \big\} \le 1 \;.
\end{align}
Without loss of generality we can assume that $\rho(t)$ is a pure state, because of the convexity of $\bTr{\rho^2}$ in $\rho$. We will show
\begin{align}
\label{PurityEquivalence}
	{\rm Tr}\big\{ [\rho(t) + d\rho_{\rm c}(t)]^2 \big\} \le 1  \; \Longleftrightarrow \;  {\sf M} {\sf M}\dg/\hbar \in \mathfrak{H} \;,
\end{align}
where ${\mathfrak H}$ is 
\begin{align}
\label{AllowedH}
	{\mathfrak H} = \left\{ {\sf H}={\rm diag}({\bm \eta}\;\!) \, | \forall \; k, \, \eta_k \in [0,1] \right\} \;,
\end{align}
thereby establishing the RHS of the equivalence \eqref{PurityEquivalence} as a necessary condition for ${\sf M}$ to be valid.

The equivalence \eqref{PurityEquivalence} can be proven by first proving the following lemma:
\begin{align}
\label{NecessaryConditonForMIdeal}
	{\rm Tr}\big\{ [\rho(t) + d\rho_{\rm c}(t)]^2 \big\} = 1  \; \Longleftrightarrow \;  \MMdg/\hbar = \ID{L} \;,
\end{align}
where $\ID{L}$ is the $L \times L$ identity matrix. It can then be shown (see Appendix~\ref{AppendixForM}) that
\begin{align}
\label{PurityInTermsOfAlpha}
	\Tr\big\{ [\rho(t) + d\rhoc(t)]^2 \big\} = 1 + \frac{2}{\hbar} \, \tr\big[ H \big( \MMdg/\hbar - \ID{L} \big)^{\!\top} \big] \, dt \;,
\end{align}
where $H$ is the Hermitian, positive-semidefinite matrix (not to be confused with the Hamiltonian $\hat{H}$, which is an operator)
\begin{align}
\label{CovarianceOfLindbladOperators}
	H = \left\langle \big( \hc - \an{\hc} \!\big)\ddg \big(\hc - \an{\hc} \!\big)^{\!\!\top} \right\rangle \;.
\end{align}
Since $\rho(t)$ is an arbitrary pure state, $H$ can be assumed to be strictly positive (i.e positive definite). Thus \eqref{PurityInTermsOfAlpha} will be 1 if and only if $\MMdg/\hbar = \ID{L}\,$. This completes the proof of \eqref{NecessaryConditonForMIdeal}, a necessary condition on ${\sf M}$ in the case of efficient monitoring (i.e. monitoring which preserves the purity of the state).

To prove \eqref{PurityEquivalence} we must consider the case when the purity at time $t+dt$ drops to below 1. Given that we had a pure state at time $t$ this is possible if and only if we had inefficient monitoring. For a single decay channel we can bundle the sum of all losses \footnote{Note that ``loss'' here refers to \emph{any} process which leads to a \emph{loss of information}, not necessarily imperfections of the measuring device.} into a single parameter $\eta$, and consider only a fraction $\eta$ (between 0 and 1) of the system operator $\hat{c}$ to be measured perfectly. This can by modelled by introducing an imaginary beam-splitter with transmission coefficient $\eta$ in the path of the decay channel followed by an ideal detector \cite{LP95}. When multiple decay channels are present we simply repeat this setup for each channel. This motivates us to rewrite the master equation \eqref{LindbladForm} as
\begin{align}
	\hbar \;\! \dot{\rho} = - i \;\! [\hat{H},\rho] + {\cal D}\big[ \rt{\ID{L}-{\sf H}} \hc \big]\rho + {\cal D}\big[ \rt{\sf H} \hc \big]\rho \;,
\end{align}
where ${\sf H} \in \mathfrak{H}$ [recall \eqref{AllowedH}], and unravel the last term with unit detection efficiency. Such an unravelling is defined by the stochastic master equation
\begin{align}
\label{NonunitDetectionSME}
	\hbar \;\! d\rhoc = {\cal L} \;\! \rho \, dt + d{\bf w}\tp {\cal H}[{\sf M'}\dg \hc'] \;\! \rho \;, 
\end{align}
where $\hc' \equiv \rt{\sf H} \hc$ and \eqref{NonunitDetectionSME} is conditioned on the current
\begin{align}
\label{NonunitDetectionCurrent}
	{\bf y}_{\sf M'} \, dt = \frac{dt}{\hbar} \, \ban{{\sf M'}\dg \hc' + {\sf M'}\tp \hc'^{\ddag}} + d{\bf w}  \;.
\end{align}
Since \eqref{NonunitDetectionSME} and \eqref{NonunitDetectionCurrent} describe efficient monitoring, from the lemma of \eqref{NecessaryConditonForMIdeal}, ${\sf M'}$ must satisfy
\begin{align}
\label{ConstraintOnM'}
	{\sf M'} {\sf M'}\dg / \hbar = \ID{L} \;.
\end{align}
The unravelling defined by \eqref{NonunitDetectionSME}, \eqref{NonunitDetectionCurrent}, and \eqref{ConstraintOnM'} is equivalent to \eqref{GeneralNonlinearDiffSME}, \eqref{GeneralDiffMmtRecord}, and \eqref{TrueyAvg2}, but the latter make the measurement efficiency ${\sf H}$ explicit. The set of quantum trajectories generated by \eqref{GeneralNonlinearDiffSME} must therefore be the same as the set generated by \eqref{NonunitDetectionSME}. This will be the case if and only if the two equations have the same stochastic term, i.e.
\begin{align}
	d{\bf w}\tp {\cal H}[{\sf M}\dg \hc] \;\! \rho = d{\bf w}\tp {\cal H}[ {\sf M'}\dg \hc'] \;\! \rho \;.
\end{align}
This is true for any $\rho$ and $\hc$ if and only if ${\sf M} = \rt{\sf H} {\sf M}'$. This gives
\begin{align}
\label{NSConditionForM}
	{\sf M} \;\! {\sf M}\dg = {\sf H} \;.
\end{align}
We have arrived at \eqref{NSConditionForM} by considering inefficient detection. However there are other properties of the measurement that one would like to capture with ${\sf M}$ so it would seem that \eqref{NSConditionForM} is necessary but not sufficient. As we show next, \eqref{NSConditionForM} is in fact, surprisingly, a sufficient constraint on ${\sf M}$ for it to be a valid parameterization of general diffusive measurements.

\subsubsection{Sufficient condition for ${\sf M}$}
\label{SuffCondForM}

To show that \eqref{NSConditionForM} is a sufficient condition we will connect ${\sf M}$ to another parameterization ${\sf U}$, which is a different way of representing diffusive measurements and has conditions that have previously been shown to be necessary and sufficient \cite{WD01,WD05}. In this paper we introduce an elegant way to connect \mrep\ and \urep\ by considering a generalized diffusion operator which we denote by $\hat{\sf D}$. If the system state is in an $N$-dimensional Hilbert space $\mathbb{H}$, then $\hat{\sf D}$ is an operator in $\mathbb{H} \otimes \mathbb{H}$, defined by
\begin{align}
\label{DiffusionOperator}
	\hat{\sf D} \, dt  \equiv  d\rhoc \otimes d\rhoc  \;.
\end{align} 
This is the mathematical object which characterizes the set of all equivalent representations (i.e. \mrep-matrices) of a given unravelling as we now explain.

Stochastic paths of the quantum state itself are rather abstract but we can make the trajectories more concrete by considering
\begin{align}
\label{VecorizedRho}
	\bm{\rho}_{\rm c} = {\rm Tr}[ \;\! \hat{\bf e} \;\! \rho_{\rm c} \;\! ]  \;, 
\end{align}
where $\hat{\bf e}=(\hat{e}_{1},\hat{e}_{2},\ldots,\hat{e}_{N^2})\tp$ is an operator-basis for all linear Hermitian operators. Thus ${\bm \rho}_{\rm c}$ is a stochastic process in $\mathbb{R}^{N^2}$, satisfying
\begin{align}
	d{\bm \rho}_{\rm c} = {\bm A} \;\! dt + B \;\! d{\bf w} \;.
\end{align}
Note that ${\bm A}=\bTr{\vop{e}\;\!{\cal L} \rho}$ is a vector while 
\begin{align}
\label{CoeffMatrixOfdw}
	B = {\rm Tr}\!\left\{ \hat{\bf e} \;\! {\cal H}[\hc\tp {\sf M}] \rho \right\} 
\end{align}
is a matrix. Recall that ${\cal L}$ and ${\cal H}[\hc\tp\mrep]$ are defined by \eqref{LindbladForm} and \eqref{VopvaluedCalH} respectively. The diffusion matrix for ${\bm \rho}_{\rm c}$ is given by 
\begin{align}
\label{DiffusionOfVecRho}
	D \;\! dt = d{\bm \rho}_{\rm c} \, d{\bm \rho}\tp_{\rm c} = B B\tp  dt \;.
\end{align}
We can see how $\hat{\sf D}$ arises by rewriting the RHS of the first equality in \eqref{DiffusionOfVecRho},
\begin{align}
	D \;\! dt = {}& {\rm Tr}\big[\hat{\bf e} \;\! d\rho_{\rm c} \big] \, {\rm Tr}\big[\hat{\bf e}\tp d\rho_{\rm c} \big]  \nn \\
\label{DasTraceDhat}
	          = {}& {\rm Tr}\big[ (\hat{\bf e} \;\! d\rho_{\rm c}) \otimes (\hat{\bf e}\tp d\rho_{\rm c}) \big] 
	          = {\rm Tr}\big[ (\hat{\bf e} \otimes \hat{\bf e}\tp) \;\! \hat{\sf D} \big] \;.
\end{align}
Thus $\hat{\sf D}$ is an operator-valued diffusion coefficient whose trace against $\hat{\bf e} \otimes \hat{\bf e}\tp$ gives us the diffusion of the more tangible process \eqref{VecorizedRho}. Turning $\rhoc$, a stochastic process in $\mathbb{H}$, into ${\bm \rho}_{\rm c}$, an equivalent stochastic process in $\mathbb{R}^{N^2}$, allows us to grasp the abstract diffusion operator $\hat{\sf D}$ by using well known properties of classical stochastic processes; namely that the diffusion of ${\bm \rho}_{\rm c}$ is characterized by the matrix $D$, and that for a given $D$ there are many matrices $B$ such that $BB\tp=D$. The equivalence between ${\bm \rho}_{\rm c}$ and $\rhoc$, and more specifically between $B$ and \mrep, and also between $D$ and $\hat{\sf D}$ (as given by \eqref{CoeffMatrixOfdw} and \eqref{DasTraceDhat}), mean that
$\hat{\sf D}$ characterizes the diffusion of $\rhoc$ just as $D$ characterizes the diffusion of ${\bm \rho}_{\rm c}$. There will be many matrices \mrep\ that generate the same $\hat{\sf D}$ (through the nonlinear term ${\cal H}[\hc\tp\mrep]\rho$) just as there are more than one $B$ corresponding to a given $D$.

We mentioned earlier in Sec.~\ref{SchPicParamUnrav} that an unravelling can be defined as the set of solutions to the stochastic master equation and this set will correspond to some ${\sf M}$. In general the set of solutions will change everytime ${\sf M}$ is changed but it may also be possible to have two different choices of ${\sf M}$, say ${\sf M}_1$ and ${\sf M}_2$ which generate the same solution set. In this case we say that ${\sf M}_1$ and ${\sf M}_2$ are equivalent representations of the same unravelling. Let us substitute \eqref{GeneralNonlinearDiffSME} into \eqref{DiffusionOperator} and define
\begin{align}
\label{DefnOfT}
	{\sf T} \equiv \tbo{{\sf T}_1}{{\sf T}_2} \equiv \tbo{\Re[{\sf M}]}{\Im[{\sf M}]}  \;,
\end{align}
where ${\sf T}_1$ and ${\sf T}_2$ are each real, and $L \times 2L$ so that ${\sf T}$ is a real $2L \times 2L$ matrix. We have defined the real and imaginary parts of an arbitrary complex matrix $A$ by 
\begin{align}
	\Re[A] \equiv \frac{1}{2} \, \big( A + A^* \big)  \;,  \quad 
	\Im[A] \equiv \frac{-i}{2} \, \big( A - A^* \big)  \;.
\end{align}
We can then write ${\sf M}$ as
\begin{align}
\label{DefnOfM}
	{\sf M} = {\sf T}_1 + i \, {\sf T}_2 \;.  
\end{align}
The vop ${\sf M}\dg \hc$ can be rewritten as
\begin{align}
\label{MTidentity}
	{\sf M}\dg \hc = {\sf T}\tp \vhc \;,   
\end{align}
where $\vhc\tp \equiv \obt{\hc\tp}{-i\;\!\hc\tp}$. Using this, \eqref{DiffusionOperator} is
\begin{align}
	\hat{\sf D} \;\! dt = {}& {\cal H}[\;\!\vhc\tp  {\sf T} \;\!d{\bf w}] \;\! \rho  
	                          \otimes  {\cal H}[d{\bf w}\tp {\sf T}\tp \vhc\;\!] \;\! \rho  \nn \\
	                    = {}& {\cal H}[\;\!\vhc\tp] \;\! \rho \, {\sf T} \;\! d{\bf w} 
	                          \otimes  d{\bf w}\tp {\sf T}\tp  {\cal H}[\vhc] \;\! \rho  \nn \\
\label{HatDInTermsOfU}
                      = {}& \hbar \, {\cal H}[\;\!\vhc\tp] \;\! \rho \, \otimes  {\sf U} \, {\cal H}[\vhc] \;\! \rho \, dt \;.
\end{align}
In the last line we have defined a $2L \times 2L$ real matrix ${\sf U}$ by
\begin{align}
\label{TTtranspose}
	{\sf T} \;\! {\sf T}\tp = \hbar \;\! {\sf U} \equiv \hbar \tbt{{\sf U}_{11}}{{\sf U}_{12}}{{\sf U}_{21}}{{\sf U}_{22}}  \;,
\end{align}
where ${\sf U}_{mn}$ are $L \times L$ blocks of ${\sf U}$. Equation \eqref{HatDInTermsOfU} tells us that given the set of Lindblad operators $\hc$, ${\sf U}$ is what uniquely specifies $\hat{\sf D}$, and therefore the unravelling. That ${\sf U}$ defines the unravelling was independently formulated in Ref.~\cite{WD01} for the case of pure-state trajectories, and later generalized to include non-unit detection efficiency in Ref.~\cite{WD05}. The necessary and sufficient conditions for a ${\sf U}$ to be valid (in the same sense as \eqref{ValidMStatement}) are
\begin{align}
\label{NSCondForU3} 
	{\sf U} \ge 0 \;, \\
\label{NSCondForU1}
	{\sf U}_{11} + {\sf U}_{22} \in \mathfrak{H}  \;,  \\
\label{NSCondForU2}
	{\sf U}_{12} = {\sf U}_{21}  \;.
\end{align}
We will denote the set of valid ${\sf U}$-matrices as 
\begin{align}
\label{DefnOfUrep}
	\mathfrak{U} = \left\{ \tbt{{\sf U}_{11}}{{\sf U}_{12}}{{\sf U}_{21}}{{\sf U}_{22}}  \, \Big| \, 
	                       {\sf U} \ge 0, {\sf U}_{11} + {\sf U}_{22} \in \mathfrak{H}, 
	                       {\sf U}_{12} = {\sf U}_{21} \right\}  \;.
\end{align} 
The question is now whether any ${\sf U}$ derived from \eqref{NSConditionForM} is valid. Since ${\sf T}$ is a real matrix, the definition \eqref{TTtranspose} ensures \eqref{NSCondForU3} for any ${\sf T}$. Using \eqref{TTtranspose} conditions \eqref{NSCondForU1} and \eqref{NSCondForU2} become, respectively,
\begin{align}
\label{NSCondForUInTermsOfT1}
	{\sf T}_1 \;\! {\sf T}\tp_1 + {\sf T}_2 {\sf T}\tp_2 \in \mathfrak{H} \;, \\
\label{NSCondForUInTermsOfT2}	
	{\sf T}_1 \;\! {\sf T}\tp_2 = {\sf T}_2 \;\! {\sf T}\tp_1 \;.
\end{align}
This defines the set of valid ${\sf T}$-matrices
\begin{align}
	\mathfrak{T} = \left\{ \tbo{{\sf T}_1}{{\sf T}_2} \, \Big| \, 
	                       {\sf T}_1 \;\! {\sf T}\tp_1 + {\sf T}_2 {\sf T}\tp_2 \in \mathfrak{H}, 
	                       {\sf T}_1 \;\! {\sf T}\tp_2 = {\sf T}_2 \;\! {\sf T}\tp_1 \right\}  \;.
\end{align}
It is easy to see, on substituting \eqref{DefnOfM} into \eqref{NSConditionForM}, that \eqref{NSCondForUInTermsOfT1} and \eqref{NSCondForUInTermsOfT2} are satisfied and thereby showing the sufficiency (and in fact necessity) of \eqref{NSConditionForM} for ${\sf M}$ to be a valid representation of an arbitrary diffusive unravelling: 
\begin{align}
	\urep(\mrep) \in {\mathfrak U} \; \Longleftrightarrow \; \mrep \in \mathfrak{M}
\end{align}
where ${\mathfrak M}$ is the set of valid ${\sf M}$-matrices
\begin{align}
\label{DefnOfMrep}
	{\mathfrak M} = \left\{ {\sf M} \, | \, {\sf M} {\sf M}\dg \in {\mathfrak H} \right\} \;.
\end{align}
Note that even though we have imported a previous result which proves altogether the sufficiency and necessity of \eqref{NSConditionForM}, our analysis in Sec.~\ref{NecessConditionForM} remains instrumental; since without it there is no reason to consider the set ${\mathfrak M}$. For ease of reference we call the members of ${\mathfrak M}$ ``M-reps'', and similarly the members of ${\mathfrak U}$ ``U-reps''.

\subsection{Comparison between ${\mathfrak M}$ and ${\mathfrak U}$}
\label{MrepsVersusUreps}

Here we discuss some general features of M-reps and compare them with U-reps. To do so we first recall the essential features of the U-reps from Refs.~\cite{WD01,WD05}. There, Wiseman and co-workers derived a stochastic master equation of the form  
\begin{align}
\label{WisemanDohertySME}
	d\rhoc = {\cal L} \;\! \rho(t) \, dt + {\cal H}[d{\bf z}\dg \hc\;\!] \;\! \rho(t)
\end{align}
(with ${\cal L}$ given by \eqref{LindbladForm}) that conditioned on an $L \times 1$ complex current
\begin{align}
\label{UrepCurrent}
	{\bf J} \, dt = \ban{{\sf H} \;\! \hc + {\sf Y} \;\! \hc\ddg} \, dt + d{\bf z}  \;,
\end{align}
where $d{\bf z}$ (also $L \times 1$ by default) is a complex white-noise increment defined by ${\rm E}[d{\bf z}]={\bf 0}$ and the correlations
\begin{align}
	d{\bf z} \, d{\bf z}\dg = \hbar \, {\sf H} \, dt \;,  \quad  d{\bf z} \, d{\bf z}\tp = \hbar \, {\sf Y} \, dt \;.
\end{align}
%
%
The matrix ${\sf H}$ is a member of $\mathfrak{H}$, while ${\sf Y}$ is complex and only has to be symmetric. The matrix \urep, or the ``unravelling matrix'' as it was originally called, is defined as the correlations of $\big( \Re[d{\bf z}\tp],\Im[d{\bf z}\tp] \big)/\rt{\hbar}$. This can be shown to be 
\begin{align}
\label{WisemanDohertyMatrix}
	\urep = \frac{1}{2} \tbt{{\sf H}+\Re[\sf Y]}{\Im[\sf Y]}{\Im[\sf Y]}{{\sf H}-\Re[\sf Y]}  \;,
\end{align}
which is consistent with \eqref{DefnOfUrep}.

The first noteworthy feature of M-reps is that they capture physically valid measurements by a single equation, \eqref{NSConditionForM}, whereas U-reps require three, namely \eqref{NSCondForU3}--\eqref{NSCondForU2}. This point makes our theory of M-reps rather elegant. If however we compare the number of real parameters required to specify ${\sf M}$ to that of ${\sf U}$, then we find that U-reps are a more efficient parameterization: We know directly from \eqref{WisemanDohertyMatrix} that the number of real parameters in ${\sf U}$ is $L^2+2L$, being the total number of real parameters required to specify ${\sf H}$ and ${\sf Y}$. To find the number of real parameters required to specify ${\sf M}$, we note that ${\sf M}$ is $L \times 2L$, so the total number of real parameters in ${\sf M}$ is $4L^2$. But \eqref{NSConditionForM} means that $L^2-L$ of these parameters are redundant (since it imposes $L^2-L$ independent real conditions on the elements of ${\sf M}$). Subtracting this redundancy from $4L^2$ leaves $3L^2+L$. Thus M-reps require $2L^2-L$ more parameters than U-reps. 

The second attractive feature of M-reps is that ${\sf M}$ can be understood as the generalization of $\rt{\sf H}$, encapsulating more information than just measurement efficiency. This was in fact part of our motivation for developing M-reps and postulating \eqref{GeneralNonlinearDiffSME}, as stated at the start of Sec.~\ref{DerivanOfMeasRecord}. So in hindsight, the fact that ${\sf M}$ turns out to be the matrix square root of ${\sf H}$, i.e. constrained by \eqref{NSConditionForM}, seems ``natural'' given that we were thinking of generalizing $\rt{\eta}$ to $\rt{\sf H}$, and then from $\rt{\sf H}$ to ${\sf M}$ from the start. The interpretation of U-reps on the other hand was derived (summarized in \eqref{WisemanDohertySME}--\eqref{WisemanDohertyMatrix}) as the correlation between the real and imaginary parts of a complex measurement output. This has the advantage that it reflects directly the quantum trajectory diffusion as was evident in \eqref{HatDInTermsOfU}. Equations \eqref{NSCondForU3}--\eqref{NSCondForU2} are simply constraints on the allowed correlations of $d{\bf z}$.

Finally, we mentioned before (two paragraphs under \eqref{MrepCurrent}) that we prefer to work with a real current, namely ${\bf y}_\mrep$, for the purpose of feedback control even though quantum measurement theory does not impose such a restriction. However, by allowing the system state to be conditioned on a real vector, ${\bf y}_{\sf M}$ now has the capacity to be an observable (i.e. correspond to an actual physical process) rather than a mere dummy variable for distinguishing the different measurement outcomes. Of course, one may also prefer a real current over a complex one simply because it is real. Motivated by exactly these reasons Ref.~\cite{WD05} defines a real current from ${\bf J}$ as
\begin{align}
\label{yUrepCurrent}
	{\bf y} \, dt \equiv \trep^+ \tbo{\Re[{\bf J}]}{\Im[{\bf J}]} dt
\end{align}
where $\trep^+$ is the Moore--Penrose inverse of \trep\ \cite{AB03}. Substituting \eqref{UrepCurrent} into \eqref{yUrepCurrent} and using the properties of $\trep^+$ it can be shown that \eqref{yUrepCurrent} simplifies to the form of \eqref{MrepCurrent} (recall \eqref{DefnOfM}). Thus the U-rep achieves the same current as the M-rep but less directly by deriving ${\bf y}$ via ${\bf J}$.

\subsection{Relations between $\mathfrak{T}$, $\mathfrak{U}$, and $\mathfrak{M}$}

The relation between $\mathfrak{M}$ and $\mathfrak{T}$ is straightforward, given by \eqref{DefnOfT} and \eqref{DefnOfM}. From these it should be clear that $\mathfrak{M}$ maps one-to-one and onto $\mathfrak{T}$. We said above that M-reps are parameterized by $3L^2+L$ real numbers, therefore ${\sf T}$ also has the same number of real parameters. This number is $2L^2-L$ more than the number of real parameters in the U-rep. We can see where the extra parameters lie by returning to \eqref{TTtranspose} which defines ${\sf T}$ as the matrix square root of ${\sf U}$. We can always write, using the polar decompostion, ${\sf T} =  \rt{{\sf T}{\sf T}\tp/\hbar} {\sf O} \equiv \rt{\sf U} {\sf O}$ where $\rt{\sf U}$ is the positive matrix square root of ${\sf U}$ \footnote{The positive square root of a matrix $A$ is defined by a matrix $B$ such that $B^2=A$, and is itself positive-semidefinite. The positive square root always exists and is uniquely given by $A$ so it is common to denote it by $\rt{A}$.} and ${\sf O}$ is some $2L \times 2L$ orthogonal matrix which may or may not be unique given \trep\ (see following paragraph). The number of parameters in $\rt{\sf U}$ is $L^2+2L$, inherited from ${\sf U}$, while the number of parameters in ${\sf O}$ is $2L^2-L$ \footnote{An $n \times n$ real orthogonal matrix is parameterized by $n(n-1)/2$ real numbers.}. Adding the number of parameters in $\rt{\sf U}$ and ${\sf O}$ gives $3L^2+L$. The extra parameters thus lie in ${\sf O}$, leading us to define the set of pairs (not to be confused with a block matrix)
\begin{align}
\label{UOrthogonalSet}
	\mathfrak{U}_{\sf O} = \big\{ \big(\rt{\sf U},{\sf O}\big) \, | \, 
	                              {\sf U} \in \mathfrak{U}, {\sf O}\tp = {\sf O}^{-1} \big\} \;,
\end{align}
where ${\sf O}$ is understood to have dimensions $2L \times 2L$.

\begin{figure}
\includegraphics[width=8.5cm]{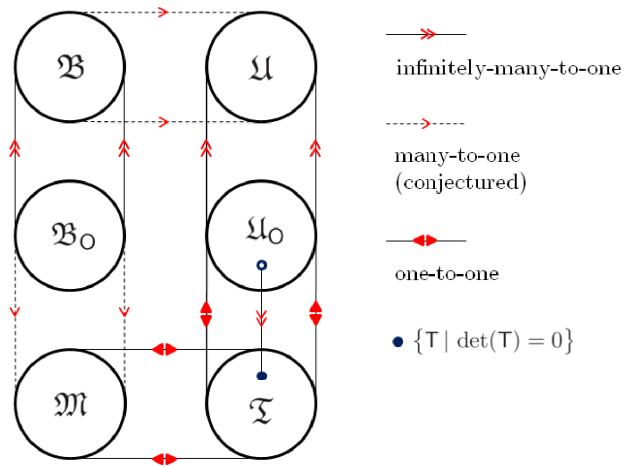} 
\caption{Multivalued mappings between the different parameterizations of diffusive quantum measurements. The lines have been drawn tangent to the circles to denote that one set is mapped onto the other, i.e. each member of one set has a corresponding member in the set to which they are connected to. Note also that we have defined a subset of $\mathfrak{T}$ whose members are non-invertible. This is a set of measure zero and maps to subset of $\mathfrak{U}_{\sf O}$ (shown as an unshaded dot), also a set of measure zero.}
\label{Parameterizations}
\end{figure}

We can divide the set $\mathfrak{T}$ into two subsets: a subset of invertible ${\sf T}$s and a subset of non-invertible ${\sf T}$s. The subset of ${\sf T}$s that are invertible maps one-to-one and onto $\mathfrak{U}_{\sf O}$, since in this case the polar decomposition of ${\sf T}$ is unique, i.e. given any ${\sf T}$, both entries of the pair $(\rt{{\sf T}{\sf T}\tp/\hbar},{\sf O}) \in \mathfrak{U}_{\sf O}$ are uniquely determined by ${\sf T}$. The subset of ${\sf T}$s that are not invertible maps infinitely-many-to-one and onto a subset of $\mathfrak{U}_{\sf O}$. In this case the first entry of the pair $(\rt{{\sf T}{\sf T}\tp/\hbar},{\sf O})$ will still be uniquely determined by ${\sf T}$ but corresponding to this ${\sf T}$ there are an infinite number of orthogonal matrices ${\sf O}$. We depict these relations in Fig.~\ref{Parameterizations} along with another parameterization $\mathfrak{B}$ to be introduced in Sec.~\ref{BlockRepresentation}. Note the cardinality of the non-invertible subset of $\mathfrak{T}$ is less than the cardinality of $\mathfrak{T}$. That is, the former is a subset of measure zero.
%

\subsection{The M-representation in the \hei\ Picture}

We now consider diffusive quantum measurements in the \hei\ picture where operators evolve and states are static, equal to their initial value. Moving from the \sch\ picture to the \hei\ picture requires two conceptual changes which we summarize below. In Sec.~\ref{BlockRepresentation} we will formulate another representation of diffusive measurements in the \hei\ picture so the discussion here apply to Sec.~\ref{BlockRepresentation} as well.

\begin{figure}
\includegraphics[width=8.5cm]{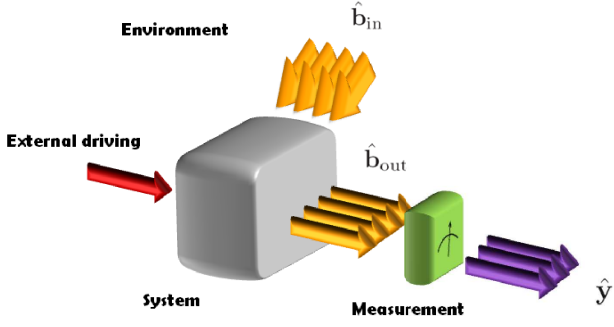} 
\caption{Quantum entanglement and measurement are two processes which have conventionally been described in the \sch\ picture. When we move to the \hei\ picture we move to a different mindset, where the system-bath entanglement and the measurement process are all captured by time-dependent operators, given by \eqref{InputOutput} \cite{Rub01,dEs95} and \eqref{SelfNondemolition}.}
\label{OverallSchemeHeiPic}
\end{figure}

The bath field $\hat{\bf b}_{\rm in}$ is now a vop-valued white-noise process. As such, $d\hat{\bf B}_{\rm in}(t) \equiv \hat{\bf b}_{\rm in}(t)\;\!dt$ satisfies \ito\ rules analogous to \eqref{ClassicalIto1}--\eqref{ClassicalIto3} \cite{Par92,GC85}. That is 
\begin{align}
	\big\langle d\hat{\bf B}_{\rm in}(t) \big\rangle = {\bf 0} \;,
\end{align}
where the angle brackets denote a quantum average and 
\begin{align}
	d\hat{\bf B}_{\rm in}(t) \, d\hat{\bf B}\dg_{\rm in}(t) = \hbar \, \mopID{L} \, dt \;,
\end{align}
with all other second-order moments negligible. Note that  
\begin{align}
	\mbrac{\dBin(t)}{\dBin\ddg(t')} = 0 \quad \forall \; t \ne t'  \;,
\end{align}
where we have used the mop-bracket \eqref{MatrixBracket}, defined for any two vops $\vop{A}$ and $\vop{B}$ as
\begin{align}
	\mbrac{\vop{A}}{\vop{B}} = \vop{A} \vop{B}\tp - \big( \vop{B} \vop{A}\tp \big)\tp  \;.
\end{align}
Thus all other second-order moments with unequal times also vanish. As a consequence of the singularity of ${\vop b}_{\rm in}$ it is necessary to distinguish the bath field before and after its interaction with the system. This gives rise to an output field (Fig.~\ref{OverallSchemeHeiPic})
\begin{align}
	\dBout(t) = \hat{U}\dg(t+dt,t) \, \dBin(t) \, \hat{U}(t+dt,t) 
\end{align}
where $\hat{U}(t+dt,t)$ is defined by \eqref{HudsonParthasarathySch} but with ${\vop b}$ replaced by $\bin$. From this one can show that
\begin{align}
\label{InputOutput}
	\dBout(t) = \hc(t) \;\! dt + \dBin(t) \;.
\end{align}
This input-output relation and is what allows us to relate our measurement performed on the bath back to properties of the system \cite{GC85,CG84}.

We next turn the measured current ${\bf y}(t)$ into a vop \cite{NCN+09,Ral99,SSH+87}. Note that this does not mean we are taking the measured current to be a quantum object. The current is interpreted as a classical quantity but it is represented by a vop in the theory. The classicality of the measured current can be formally captured by the ``self-nondemolition'' property of the vop $\vop{y}(t)$ \cite{Bel89,BS89,BHH95},
\begin{align}
\label{SelfNondemolition}
	\mbrac{\vop{y}(t)}{\vop{y}(t')} = 0 \;, \;  \forall \; t,t' \;.
\end{align} 
It should also correspond to an observable process so $\vop{y}(t)$ must be Hermitian. In brief, quantum measurement theory in the \hei\ picture just assigns an operator $\vop{y}(t)$ (a quantum stochastic process) which reproduces the correct measurement statistics (i.e. the statistics of ${\bf y}(t)$, a classical stochastic process) \cite{WM10}.

If a vop $\vop{y}$ is to represent the measured current ${\bf y}$ then it must be such that the statistics of ${\bf y}$ are reproduced, namely,
\begin{align}
\label{MeanOfyHat}
	\ban{\vop{y}(t)} = {\rm E}\big[ {\bf y}(t) \big]   \;,
\end{align}
and
\begin{align}
\label{VarianceOfyHat}
	\big[ \vop{y}(t) \, dt \big]  \big[ \vop{y}(t) \, dt \big]\tp 
	= \big[ {\bf y}(t) \, dt \big] \big[ {\bf y}(t) \, dt \big]\tp \hat{1}  \;.
\end{align}
Physically the measurement is never performed directly on the system but rather on the bath which involves detecting the output fields. Therefore $\vop{y}$ should depend on $\bout$. It thus makes sense to write 
\begin{align}
\label{MrepVop}
	\hbar \, \vop{y}_{\sf M} \, dt = {\sf M}\dg \dBout + {\sf M}\tp \dBout\ddg + \hbar \, \dups_{\sf M} \;.
\end{align}
We have defined the noise increment $\dups_\mrep$ to be
\begin{align}
	\dups_{\sf M} = {\sf L} \, d\vop{U} + {\sf L}^{\!*} \, d\vop{U}\ddg  \;,
\end{align}
where ${\sf L}$ is $2L \times 2L$, to be chosen so that \eqref{MeanOfyHat} and \eqref{VarianceOfyHat} are satisfied. The vop $d\vop{U}$ is a free vacuum field just like $\dBin$ except that it is $2L \times 1$ so the only non-vanishing moment is
\begin{align}
	d\vop{U}(t) \, d\vop{U}\dg(t) = \hbar \, \mopID{2L} \, dt  \;.
\end{align}
Note that this means \eqref{MeanOfyHat} is automatically satisfied. Enforcing \eqref{VarianceOfyHat} we obtain
\begin{align}
\label{ConditionForV}
	\hbar^2 \, {\sf L} \, {\sf L}\dg = \hbar \, \ID{2L} - \MdgM   \;.
\end{align}
For convenience we will define ${\sf Z} = \hbar \, \ID{2L} - \MdgM $. We can thus choose $\hbar \, {\sf L}$ to be the positive square root of the RHS of \eqref{ConditionForV} and write
\begin{align}
	\hbar \, {\sf L} = \rt{\sf Z}  \;.
\end{align}
Note that $\dups_{\sf M}$ is not a Wiener increment (it has correlations given by ${\sf Z}\;\!dt/\hbar$) but we can write $\vop{y}_{\sf M}$ as
\begin{align}
\label{MrepVop2}
	\hbar \, \vop{y}_{\sf M} \, dt = \big( {\sf M}\dg \hc + {\sf M}\tp \hc\ddg \big) dt + \hbar \, [d\vop{v}_{\rm m}]_\mrep
\end{align}
where
\begin{align}
	\hbar \, [d\vop{v}_{\rm m}]_\mrep = {\sf M}\dg \dBin + {\sf M}\tp \dBin\ddg + \hbar \, {\sf L} \, d\vop{U} + \hbar \, {\sf L}^{\!*} \, d\vop{U}\ddg
\end{align}
is a Wiener increment (when divided by $\hbar$), i.e.
\begin{align}
\label{QuantumWienerReal1}
 	 [d\vop{v}_{\rm m}(t)]_\mrep \, [d\vop{v}_{\rm m}(t)]\tp_\mrep = {}& \ID{2L} \, dt  \;,  \\
\label{QuantumWienerReal2}
	[d\vop{v}_{\rm m}(t)]_\mrep \, [d\vop{v}_{\rm m}(t')]\tp_\mrep = {}& 0  \quad \forall \; t \ne t' \;.
\end{align}
We can show that $\vop{y}_{\sf M}$ satisfies the self-nondemolition property \eqref{SelfNondemolition} by first substituting either \eqref{MrepVop}, or \eqref{MrepVop2} into \eqref{SelfNondemolition} and then considering the two cases $t=t'$ and $t \ne t'$ separately. It will be quicker to use \eqref{MrepVop2} and write \eqref{SelfNondemolition} in terms vacuum increments. Doing so gives
\begin{align}
	\mbrac{\vop{y}(t)\;\!dt}{\vop{y}(t')\;\!dt} = \mbrac{\,\;\![d\vop{v}_{\rm m}(t)]_\mrep}{[d\vop{v}_{\rm m}(t')]_\mrep\,} = 0  \;,
\end{align}
by virtue of \eqref{QuantumWienerReal1} and \eqref{QuantumWienerReal2}.

Often one would measure, in the laboratory, the autocorrelation function of the photocurrent and its spectrum (which is the Fourier transform of the autocorrelation function at steady-state), hence we now consider the autocorrelation function of $\vop{y}_{\mrep}$. This is a two-time average which can be calculated by using the vop quantum regression formula \eqref{VecOpRegress1}. The derivation of the autocorrelation of $\vop{y}_{\mrep}$ follows the same method as in Ref.~\cite{CW11b} which derives the same quantity but with feedback. The only difference here is that these feedback terms would not appear so we will refer the reader to Ref.~\cite{CW11b} for details. However we make the following heuristic argument which leads us to the correct form for the correlation function: First generalizing the correlation function of the homodyne current when $L=1$ found previously in Ref.~\cite{WM93} to the case of arbitrary $L$ we obtain,
\begin{align}
\label{LHomodyneCorrFunction}
	{}& \hspace{-0.5cm} \ban{\vop{y}_{\mrep}(t) \, \vop{y}\tp_{\mrep}(t+\tau)}  \nn \\
	   = {}& \Big( {\rm Tr}\Big\{ \big(\rt{\sf H} \hc + \rt{\sf H} \hc\ddg \big)  \nn \\
	       & \times e^{{\cal L}\tau} \big[  \hc\tp \rt{\sf H} \, \rho(t) + \rho(t) \, \hc\dg \rt{\sf H} \big] \Big\} \Big)\tp 
	         + \, \hbar^2 \;\! \ID{2L} \, \delta(\tau) \;,
\end{align}
where ${\sf H} \in \mathfrak{H}$ was defined in \eqref{AllowedH} and ${\cal L}$ is given by \eqref{LindbladForm}. The action of the superoperator and the trace are defined in \eqref{SuperopMultiplication} and \eqref{MatrixOpTrace} respectively. We said in Sec.~\ref{MrepsVersusUreps} that \mrep\ is a generalization of $\rt{\sf H}$. So the next logical step would be to replace $\rt{\sf H}$ by \mrep. However this would not keep the Hermiticity of the terms standing on either side of $\exp({\cal L}\tau)$, so the correct replacements in \eqref{LHomodyneCorrFunction} should be
\begin{align}
	\rt{\sf H} \hc \; \longrightarrow \; \mrep\dg \hc \;,  \quad  \rt{\sf H} \hc\ddg  \; \longrightarrow \; \mrep\tp \hc\ddg  \;.
\end{align}
We then obtain
\begin{align}
	{}& \hspace{-0.5cm} \hbar^2 \ban{\vop{y}_{\mrep}(t) \, \vop{y}\tp_{\mrep}(t+\tau)}  \nn \\
	 = & \Bigg( \Tr \Big\{ \big( \mrep\dg \hc + \mrep\tp \hc\ddg \big)  
	      e^{{\cal L}\tau} \Big[ \hc\tp \mrep^* \rho(t) + \rho(t) \hc\dg \mrep  \Big] \Big\} \Bigg)^{\!\!\top}  \nn \\
	   & + \, \hbar^2 \;\! \ID{2L} \, \delta(\tau) \;,
\end{align}
This is a generalization of the homodyne autocorrelation function found in Ref.~\cite{WM93} to arbitrary diffusive measurements.

\section{The Block-representation}
\label{BlockRepresentation}

\begin{figure*}
\includegraphics[width=15.0cm]{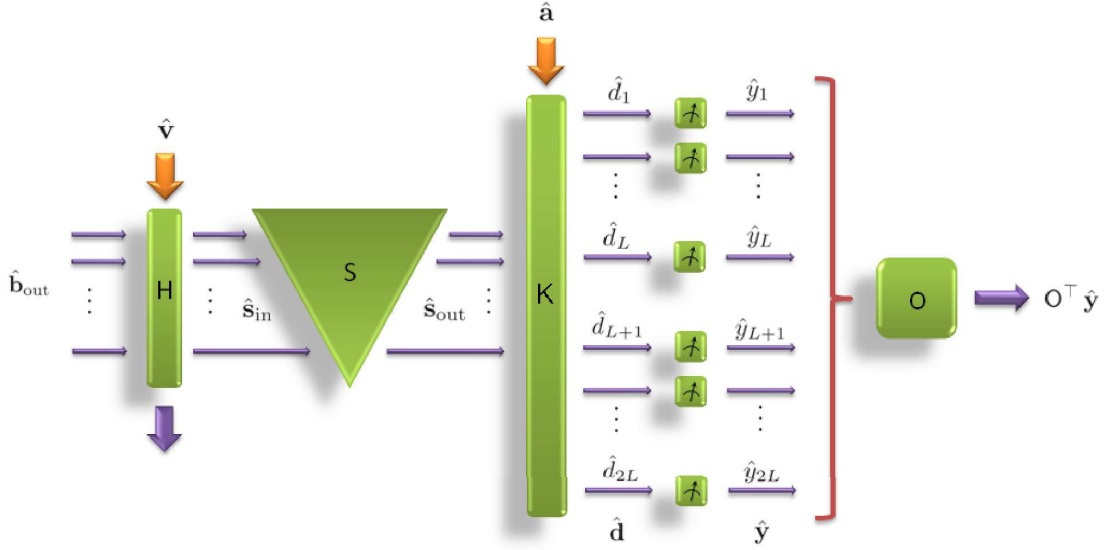} 
\caption{Decomposition of the measurement box in Fig.~\ref{OverallSchemeHeiPic} into blocks. Vacuum inputs are denoted by orange arrows. We are ignoring half of the output modes of ${\sf H}$. Note that ${\sf H}$ and ${\sf K}$ do not mix the input modes (not counting vacuum inputs) i.e. each output mode of either ${\sf H}$ or ${\sf K}$ will not depend on more than one signal mode (mode coming from the left) and one vacuum mode (mode coming from the top). We represent this in the diagram by using highly asymmetric rectangulars. Each output of ${\sf K}$ then enters a homodyne detector, set to measure the input quadrature as defined in \eqref{BlockRepCurrent} (also see Fig.~\ref{K-Block}).}
\label{BlockRep}
\end{figure*}

In this section we parameterize diffusive measurements by starting in the \hei\ picture. The advantage of this is that it allows us to model the measurement by thinking directly about the physical transformations $\hat{\bf b}_{\rm out}$ must undergo to produce a suitable current $\hat{\bf y}$. We show that the measurement block in Fig.~\ref{OverallSchemeHeiPic} can be decomposed in a way shown in Fig.~\ref{BlockRep}.

\subsection{Definition}
\label{HeisenbergCurrentOperator}

The blocks of Fig.~\ref{BlockRep} are as follows:

1) Perhaps the first property of measurements that comes to one's mind is detection inefficiencies. For a single output this can be modelled by placing a beam-splitter with transmission coefficient $\eta$ in the path of $\hat{b}_{\rm out}$. The other input to the beam-splitter is not excited (i.e. no photons) so it is simply a field $\hat{v}$ in the vacuum state. We will choose a phase convention for the beam-splitter such that its output is
\begin{align}
\label{SingleOutputBeamsplitterPhase}
	\hat{s}_{\rm in} = \rt{\eta} \, \hat{b}_{\rm out} + \rt{\bar{\eta}} \, \hat{v} \;.
\end{align}
where $\bar{\eta} \equiv 1-\eta$. Note that we are retaining only one of the beam-splitter outputs since this is all that is necessary to model detector efficiencies. When multiple outputs from the system are present \eqref{SingleOutputBeamsplitterPhase} simply generalizes to 
\begin{align}
	{\vop s}_{\rm in} = \rt{\sf H} \, {\vop b}_{\rm out} + \rt{\bar{\sf H}} \, \vop{v} \;,
\end{align}
where $\bar{\sf H}\equiv \ID{L}-{\sf H}$. This is the first block in Fig.~\ref{BlockRep}. 

2) Next, one is free to mix and phase shift the different components of $\vop{b}_{\rm out}$ to obtain an arbitrary linear combination of them. Hence we multiply $\vop{s}_{\rm in}$ by an $L \times L$ matrix to get
\begin{align}
	\vop{s}_{\rm out} = {\sf S} \, \vop{s}_{\rm in} \;.
\end{align}
We assume this mixing of the components of $\bout$ conserves total boson number so that $\vop{s}\dg_{\rm in} \vop{s}_{\rm in} = \vop{s}\dg_{\rm out} \vop{s}_{\rm out}$. This constrains ${\sf S}$ to be a unitary matrix. Note that ${\sf S}$ can be explicitly constructed out of $L(L-1)/2$ beam-splitters, $L(L+1)/2$ phase-shifters in a triangular arrangement \cite{RZB+94} as shown schematically in Fig.~\ref{BlockRep}. This scheme assumes that each component of $\bout$ have the same polarization, transverse mode structure, and mean frequency. If this were not the case then the appropriate corrctions would have to be made prior to entering the array in Fig.~\ref{BlockRep}.

\begin{figure}
\includegraphics[width=8.6cm]{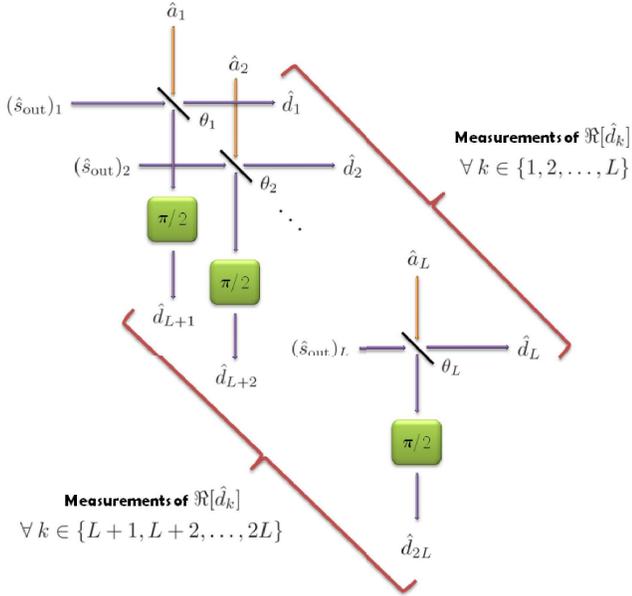}
\caption{Expanded diagram of ${\sf K}$. This prepares the components of ${\vop s}_{\rm out}$ into quadratures, ready to be measured by units of homodyne detectors.}
\label{K-Block}
\end{figure}

3) We then split each component of ${\vop s}_{\rm out}$ into quadratures. The process represented by ${\sf K}$ in Fig.~\ref{BlockRep} is shown in detail in Fig.~\ref{K-Block}. In general the intensities of the quadratures need not be equal so each beam-splitter in Fig.~\ref{K-Block} has a different transmission coefficient which we denote by $\theta_k$ ($k=1,2,\ldots,L$). This gives
\begin{align}
	\vop{d} = {\sf K} \tbo{{\vop s}_{\rm out}}{\vop a} \;.
\end{align}
If we define ${\sf Q}={\rm diag}(\bm \theta)$ where $\theta_k \in [0,1]$ for every $k$, and $\bar{\sf Q}=\ID{L}-{\sf Q}$, then ${\sf K}$ can be written explicitly as
\begin{align}
	{\sf K} = \tbt{\rt{\sf Q}}{\rt{\bar{\sf Q}}}{i\rt{\bar{\sf Q}}}{-i\rt{\sf Q}} \;.
\end{align}
Note that ${\sf K}$ is also a unitary matrix since $\vop{d}\dg \vop{d} = \vop{s}\dg_{\rm out} \vop{s}_{\rm out} + \vop{a}\dg \vop{a}$.

4) Individual homodyne detectors are then placed in the path of each component of $\vop{d}$. The detectors are ``tuned'' to give a current that is normalized to the local oscillator amplitude that is inside it. Each homodyne detector is set to measure ${\rm Re}[\hat{d}_k]$. This final current can therefore be defined to be 
\begin{align}
\label{BlockRepCurrent}
	\vop{y} = \frac{\vop{d} + \vop{d}^{\ddag}}{\rt{\hbar}} \;.
\end{align}
We see then, that once $\bm{\eta}$, $\bm{\theta}$, and ${\sf S}$ are specified, it seems plausible that steps one to four above, by construction, will simulate an arbitrary diffusive measurement. Grouping these parameters into a triple ${\sf B} \equiv (\bm{\eta},{\sf S},\bm{\theta})$ we define
the block-representation (or B-rep for short) of diffusive measurements by
\begin{align}
\label{ValidBlockReps}
	{\mathfrak B} 
	= \left\{ (\bm{\eta},{\sf S},\bm{\theta}) \, | \, \forall \,k,\; \eta_{k}, \theta_{k} \in [0,1], \; {\sf S}\dg = {\sf S}^{-1} \right\} \;.
\end{align}

5) We shall henceforth label the B-rep current in \eqref{BlockRepCurrent} as $\vop{y}_\brep$. Once $\vop{y}_\brep$ has been obtained one is free to do some post-processsing on this current. This will only be so if ${\sf O}$ is an orthogonal matrix and the post-processing produces ${\sf O}\tp \vop{y}_\brep$. We will have more to say about this in Sec.~\ref{RelationsBetweenBMU}.

\subsection{The Current Vector-operator}
\label{MeasurementRecordOperator}

Working out the input-output relation for each block in Fig.~\ref{BlockRep} we arrive at an equation for $\vop{y}$ that is a function of $\brep=(\bm{\eta},{\sf S},\bm{\theta})$. Let us denote this dependence by using a subscript on $\vop{y}$
\begin{align}
\label{BrepVop}
	\hbar \, \vop{y}_{\sf B} \, dt = {}& {\sf J} \, \dBout + {\sf J}^* \, \dBout\ddg + \hbar \, d\vog{\upsilon}_{\sf B} \;,
\end{align}
where
\begin{align}
\label{NoiseCoeffJ}
	{\sf J} \equiv \rt{\hbar} \tbo{\rt{\sf Q} \, {\sf S} \rt{\sf H}}{i\rt{\bar{\sf Q}} \, {\sf S} \rt{\sf H}}  \;,
\end{align}
and $\hbar \, d\vog{\upsilon}_{\sf B}$ is given by
\begin{align}
\label{MmtNoiseIncrement}
	\hbar \, d\vog{\upsilon}_{\sf B} = {\sf V} \, d\vop{V} + {\sf V}^* \, d\vop{V}\ddg + {\sf A} \, d\vop{A} + {\sf A}^* \, d\vop{A}\ddg \;.
\end{align}
The noise increments $d\vop{V}$ and $d\vop{A}$ are mutually uncorrelated quantum Wiener increments. The coefficient matrices in \eqref{MmtNoiseIncrement} are defined by
\begin{align}
\label{NoiseCoeffV}
	{\sf V} = {}&  \frac{1}{\rt{\hbar}} \tbo{\rt{\sf Q} \, {\sf S} \rt{\bar{\sf H}}}{i \rt{\bar{\sf Q}} \, {\sf S} \rt{\bar{\sf H}}} \;, \\
\label{NoiseCoeffA}
	{\sf A} = {}& \frac{1}{\rt{\hbar}} \tbo{\rt{\bar{\sf Q}}}{-i\rt{\sf Q}} \;.
\end{align}
Like the current in the M-rep, $\dups_{\sf B}$ represents correlated noise but we can write 
\begin{align}
	\hbar \, \vop{y}_{\sf B} \, dt = {}& {\sf J} \, \hc + {\sf J}^* \, \hc\ddg + [d\vop{v}_{\rm m}]_\brep
\end{align}
where 
\begin{align}
	[d\vop{v}_{\rm m}]_\brep = {}& {\sf J} \, \dBin + {\sf J}^* \, \dBin\ddg  \nn \\
	                             & + {\sf V} \, d\vop{V} + {\sf V}^* \, d\vop{V}\ddg + {\sf A} \, d\vop{A} + {\sf A}^* \, d\vop{A}\ddg
\end{align}
has a mean of zero  
\begin{align}
\label{MeasurementQuantumNoise1}
	\ban{ [d{\vop v}_{\rm m}(t)]_\brep } = {}& {\bf 0} \;, 
\end{align}
and the only non-vanishing second-order moment
\begin{align}
\label{MeasurementQuantumNoise2}
	[d{\vop v}_{\rm m}(t)]_\brep \: [d{\vop v}_{\rm m}(t)]\tp_\brep = \mopID{2L} dt \;. 
\end{align}
These can be understood physically by considering a vacuum input, i.e. $\bout=\bin$. None of the transformations applied to the vacuum inputs in Fig.~\ref{BlockRep} alter their statistics, only add one vacuum field to another, which is again vacuum. We can also verify this intuition mathematically by using \eqref{NoiseCoeffJ}, \eqref{NoiseCoeffA}, and \eqref{NoiseCoeffV} to derive \eqref{MeasurementQuantumNoise2}. All vanishing statistics should be obvious from the properties of the vacuum inputs. One may also verify that \eqref{SelfNondemolition} is satisfied by $\vop{y}_{\sf B}$.

\subsubsection{Example --- dual-homodyne detection}
\label{DualHomodyneExample}

To illustrate how the B-rep works we will consider the familiar example of ideal heterodyne detection from quantum optics in the case of $L=1$, for which ${\sf H}\equiv\eta=1$, ${\sf S} \equiv e^{i\phi}$, and ${\sf Q}\equiv\theta=1/2$. Since the measurement has unit detection efficiency, there is no vacuum field $\hat{v}$.
%
%
It is simple to see that
\begin{align}
	{\sf K} = \frac{1}{\rt{2}} \tbt{1}{1}{i}{-i}  \;,
\end{align}
which transforms $\hat{s}_{\rm out} =e^{i\phi} \hat{b}_{\rm out}$ and $\hat{a}$ into $\vop{d}=(\hat{d}_1,\hat{d}_2)\tp$. From $\vop{d}$ the final output $\vop{y}_{\sf B}$ can be seen to be
\begin{align}
	\vop{y}_{\sf B} = \tbo{\hat{y}_{1}}{\hat{y}_{2}} 
	                = \frac{2}{\rt{2\hbar}} \tbo{\hat{X}_{\rm out}(\phi) + \Re[\hat{a}]}{\hat{X}_{\rm out}(\phi+\pi/2)+\Im[\hat{a}]} \;,
\end{align}
where we have defined the the quadrature operator
\begin{align}
	\hat{X}_{\rm out}(\phi) = ( \hat{b}_{\rm out} \, e^{i\phi} + \hat{b}\dg_{\rm out} \, e^{-i\phi} )/2 \;.
\end{align}

\subsection{Relations between $\mathfrak{B}$, $\mathfrak{M}$, and $\mathfrak{U}$}
\label{RelationsBetweenBMU}

Now that we have the M-rep, B-rep, and also the U-rep found previously in Refs.~\cite{WD01,WD05}, it would be instructive to study how one may convert from one representation to another. We can first derive a map $\mathfrak{B} \to \mathfrak{M}$ by noting that if the B-rep is to realize an arbitrary diffusive measurement then it must be the case that
\begin{align}
\label{BAvgEqualsMAvg}
	\ban{\vop{y}_{\sf B}} = \ban{\vop{y}_{\sf M}}  \;.
\end{align}
Note that this is all we need since the second-order moments have already been made equal. From \eqref{MrepVop}, \eqref{BrepVop}, and \eqref{NoiseCoeffJ} we obtain
\begin{align}
\label{MgivenB}
	{\sf M} = {\sf J}\dg = \rt{\hbar \;\! {\sf H}} \;\! {\sf S}\dg \obt{\rt{\sf Q}}{-i \rt{\bar{\sf Q}}}  \;.
\end{align}
One can verify from this that ${\sf M}{\sf M}\dg=\hbar\;\!{\sf H}$, as required if ${\sf M}$ is to be a valid representation of the measurement. Taking the real and imaginary parts of \eqref{MgivenB} we get,
\begin{align}
	{\sf T} = \rt{\hbar} \tbt{\rt{\sf H} \;\! \Re[{\sf S}\tp] \rt{\sf Q}}{-\rt{\sf H} \Im[{\sf S}\tp] \rt{\bar{\sf Q}}}
	                         {-\rt{\sf H} \;\! \Im[{\sf S}\tp] \rt{\sf Q}}{-\rt{\bar{\sf Q}} \Im[{\sf S}\tp] \rt{\sf H}}  \;.
\end{align}
From ${\sf T}$ we can construct ${\sf U}$ using the definition \eqref{TTtranspose}. This gives
\begin{align}
\label{U11givenB}
	{\sf U}_{11} = {}& \rt{\sf H} \, \Re[{\sf S}\tp]  {\sf Q} \, \Re[{\sf S}] \, \rt{\sf H}  
	                   + \rt{\sf H} \, \Im[{\sf S}\tp]  \bar{\sf Q} \, \Im[{\sf S}] \, \rt{\sf H} \;, \\
\label{U12givenB}
	{\sf U}_{12} = {}& - \rt{\sf H} \, \Re[{\sf S}\tp]  {\sf Q} \, \Im[{\sf S}] \, \rt{\sf H} 
	                   + \rt{\sf H} \, \Im[{\sf S}\tp]  \bar{\sf Q} \, \Re[{\sf S}] \, \rt{\sf H} \;, \\
\label{U21givenB}
	{\sf U}_{21} = {}& - \rt{\sf H} \, \Im[{\sf S}\tp]  {\sf Q} \, \Re[{\sf S}] \, \rt{\sf H} 
	                   + \rt{\sf H} \, \Re[{\sf S}\tp]  \bar{\sf Q} \, \Im[{\sf S}] \, \rt{\sf H} \;, \\
\label{U22givenB}
	{\sf U}_{22} = {}& \rt{\sf H} \, \Im[{\sf S}\tp]  {\sf Q} \, \Im[{\sf S}] \, \rt{\sf H}  
	                   + \rt{\sf H} \, \Re[{\sf S}\tp]  \bar{\sf Q} \, \Re[{\sf S}] \, \rt{\sf H} \;.
\end{align}
By construction this ${\sf U}$ is positive-semidefinite. By noting that if ${\sf S}$ is unitary,
\begin{align}
	\Re[{\sf S}\tp] \, \Re[{\sf S}] + \Im[{\sf S}\tp] \, \Im[{\sf S}] = \ID{L} \;,  \\
	\Re[{\sf S}\tp] \, \Im[{\sf S}] = \Im[{\sf S}\tp] \, \Re[{\sf S}]  \;,
\end{align}
we can verify that the remaining conditions for ${\sf U}$ to be a valid parameterization, \eqref{NSCondForU1} and \eqref{NSCondForU2}, are indeed satisfied by \eqref{U11givenB}--\eqref{U22givenB}.

Equation \eqref{MgivenB} says that given a \brep\ we can always find an \mrep. But if we are given a valid ${\sf M}$ can we always find a B-rep? To answer this question we begin by counting the number of parameters required to specify \brep. The total number of real parameters in ${\sf B}$ is simply the sum of the parameters in each block. It should be clear that ${\sf H}$ and ${\sf K}$ each contribute $L$ real parameters. The number of real parameters in an $L \times L$ unitary matrix is $L^2$ so the B-rep has $L^2+2L$ real parameters, the same as the U-rep. We said earlier that M-reps require $3L^2+L$ real parameters, which is $2L^2-L$ more than the B-rep implying that \eqref{MgivenB} is one-to-many from $\mathfrak{B}$ to $\mathfrak{M}$. However, observe that for any $2L \times 2L$ orthogonal matrix ${\sf O}$
\begin{align}
\label{MOrthogonal}
	{\sf M} = \rt{\hbar \;\! {\sf H}} \;\! {\sf S}\dg \obt{\rt{\sf Q}}{-i \rt{\bar{\sf Q}}} {\sf O}
\end{align}
is not only a valid M-rep but its RHS has $3L^2+L$ number of real parameters, exactly the number of parameters required to specify ${\sf M}$. It is then very natural to ask whether every valid M-rep can be factorized in the form of \eqref{MOrthogonal}. There does not seem to be a simple proof in general (for any $L$) but we do analyze the simplest ($L=1$) instance of \eqref{MOrthogonal} in Appendix~\ref{FactorizationExample}. In summary our analysis shows that for $L=1$ and for any \mrep\ we can find a $(\brep,{\sf O})$ such that \eqref{MOrthogonal} is satisfied. However it involves the solutions of transcedental equations. The appearance of these transcedental equations, even in the simplest case, suggests why proving \eqref{MOrthogonal} in general may be nontrivial. But based on the parameter-count and the verification of the $L=1$ case we conjecture that every $\mrep \in \mathfrak{M}$ has a matrix decomposition in the form of \eqref{MOrthogonal}. This motivates us to define  
\begin{align}
\label{BlockRepWithO}
	\mathfrak{B}_{\sf O} = \big\{ ({\sf B},{\sf O}) \, | \, {\sf B} \in \mathfrak{B}, {\sf O}\tp={\sf O}^{-1} \big\}  \;.
\end{align}
Equation \eqref{MOrthogonal} is then shown in Fig.~\ref{Parameterizations} by the dotted line. This is a many-to-one map from $\mathfrak{B}_{\sf O}$ to $\mathfrak{M}$ as there will be more than one orthogonal matrix ${\sf O}$ that will factorize \mrep. This can be seen directly by considering the $L=1$ case which we have analysed in Appendix~\ref{FactorizationExample}. From Fig.~\ref{ThetaAndFPlots} we can see that given an \mrep\ (and therefore $r$ and $\delta$) there is a range of values for $\varphi$ (which in turn determine ${\sf O}$ via \eqref{2X2O}) at one's disposal for factoring \mrep\ in the form of \eqref{MOrthogonal}.
%
%

We can connect the orthogonal matrix in \eqref{MOrthogonal} and \eqref{BlockRepWithO} to Fig.~\ref{BlockRep}. Since given an $\mrep \in \mathfrak{M}$ we can write $\mrep'=\mrep {\sf O}$ which is also in $\mathfrak{M}$, \eqref{MrepVop} and \eqref{BAvgEqualsMAvg} imply that
\begin{align}
	\ban{\vop{y}_{\mrep'}} = \ban{\vop{y}_{\mrep {\sf O}}} = {\sf O}\tp \ban{\vop{y}_{\mrep}} = {\sf O}\tp \ban{\vop{y}_{\brep}}  \;.
\end{align}
That is to say, if $\vop{y}_\brep$ emulates $\vop{y}_\mrep$, then ${\sf O}\tp \vop{y}_\brep$ (obtained from post-processing $\vop{y}_\brep$) will emulate $\vop{y}_{\mrep {\sf O}}$.

The completeness of the B-rep also means that every valid \urep\ must have a matrix decomposition of the form given in \eqref{U11givenB}--\eqref{U22givenB}. The map $\mathfrak{B} \to \mathfrak{U}$ is thus onto. Even though \brep\ and \urep\ are parameterized by the same number of real parameters the map $\mathfrak{B} \to \mathfrak{U}$ is actually many-to-one. To see this we can consider \eqref{U11givenB}--\eqref{U22givenB} for $L=1$. In this case \urep\ is $2 \times 2$ and ${\sf S} \equiv e^{i\phi}$, ${\sf H} \equiv \eta$, and ${\sf Q} \equiv \theta$ as in Sec.~\ref{DualHomodyneExample}. It is sufficient to consider $\eta=1$, in which case \eqref{U11givenB}--\eqref{U22givenB} simplify to
\begin{align}
	\urep_{11} = {}& \theta \, \cos^2\phi + (1-\theta) \, \sin^2\phi  \;,  \\ 
	\urep_{12} = {}& (1 -2\theta) \, \cos\phi \, \sin\phi   \;, \\
	\urep_{22} = {}& \theta \, \sin^2\phi + (1-\theta) \, \cos^2\phi   \;,
\end{align}
where $\urep_{12}$ is identically equal to $\urep_{21}$ as can be verified directly from \eqref{U12givenB} and \eqref{U21givenB} for any $L$. Remembering that $\cos(\phi+\pi)=-\cos\phi$ and $\sin(\phi+\pi)=-\sin\phi$, we can see that $\brep_1 \equiv (1,e^{i\phi},\theta)$ and $\brep_2 \equiv (1,e^{i(\phi+\pi)},\theta)$ (recall the order of the B-rep triple from \eqref{ValidBlockReps}) both map to the same \urep\ but yet $\brep_1 \ne \brep_2$. We can consider another example in which $\brep_3 \equiv (1,e^{i(\phi+\pi/2)},1-\theta)$ and $\brep_1$ map to the same \urep\ but $\brep_1 \ne \brep_3$ (this time noting $\cos(\phi+\pi/2)=-\sin\phi$ and $\sin(\phi+\pi/2)=\cos\phi$). The relation between $\mathfrak{B}$ and $\mathfrak{U}$ is shown in Fig.~\ref{Parameterizations}.


\section{Concluding Remarks}
\label{ParameterizationConclusion}

\subsection{Summary of Key Results}

In this paper we have put forth two new parameterizations of diffusive quantum measurements. The first is the M-rep introduced in Sec.~\ref{M-Representation}, defined by \eqref{DefnOfMrep}. This is a characterization of diffusive measurements in the \sch\ picture, and as a consequence is a rather abstract parameterization. However it embodies all diffusive measurements by a single equation which relates simply to the efficiency with which each output field is measured $\rt{\sf H}$. The M-rep has been shown to be a complete parameterization --- every diffusive quantum measurement can always be associated with an $\mrep \in \mathfrak{M}$, and every $\mrep \in \mathfrak{M}$ must correspond to a diffusive measurement.

We then turned to the \hei\ picture in Sec.~\ref{BlockRepresentation} and proposed a realization for arbitrary diffusive measurements. This gave another parameterization which we have called the B-rep (or block-rep), defined by \eqref{ValidBlockReps}. By construction every $\brep \in \mathfrak{B}$ must correspond to a diffusive measurement. In particular we derived \eqref{MOrthogonal} and used it to investigate the existence of B-reps for arbitrary diffusive measurements. An analysis of \eqref{MOrthogonal} for $L=1$ reveals that it requires solutions to transcedental equations, which do indeed exist, for any \mrep. The occurrence of transcedental equations even in this simplest case suggests that proving \eqref{MOrthogonal} for any $L$ may be non-trivial, or even formidable. The verification of \eqref{MOrthogonal} and the fact that this equation \eqref{MOrthogonal} has the same number of parameters on each side prompts us to propose it as a general matrix decomposition for every $\mrep \in \mathfrak{M}$, and thereby conjecturing the completeness of the B-rep.

Finally, we have explained the relations between the different parameterizations (including the U-rep) and Fig.~\ref{Parameterizations} provides a summary of these relations.

\subsection{Future Prospects}

Quantum measurements are an integral part of many quantum computing architectures. The monitoring of qubits (or qunats \cite{KL10}) and their unwanted environmental coupling provides a natural setting to apply quantum-trajectory theory. Indeed, the applicability of quantum trajectories to quantum information was recognized in 1995 by Schack and colleagues \cite{SBP95}. At the same time a second group (Pellizzari \emph{et al.}) had already \emph{applied} quantum trajectories to study decoherence in the cavity QED architecture of quantum computing \cite{PGC+95}. Subsequently quantum-trajectory theory was adopted to study the effects of decoherence on the quantum information processing of discrete-level systems to some generality in Ref.~\cite{CBC+04}, and more specifically in quantum error correction \cite{CBC03,AWM03,vHM05,Mab09}, gate performance \cite{GI05}, and teleportation (of both discrete and continuous variables) \cite{BBS+97, WC09, Car05}.

As we scale up our quantum processors one would expect quantum trajectories to remain useful. A scale-up of existing protocols would inevitably mean a scale-up of the necessary measurement from being a few-port device to many-port ones. In fact, the error syndrome which conditions the corrective displacement of just one qunat in the nine-qunat error correction scheme of Braunstein \cite{Bra98,ATK+09} already requires eight homodyne detectors. If one wanted to simulate the conditioned evolution based on this eight-homodyne output using \eqref{GeneralNonlinearDiffSME}, then it is necessary to parameterize this measurement as an \mrep, in this case of dimensions of $8 \times 16$. If all we knew was $\mrep \mrep\dg / \hbar \in \mathfrak{H}$ then it is not so easy to see what \mrep\ should be, at least not without some thought. Parameterizing this measurement into a B-rep on the other hand is much easier. One could then use \eqref{MOrthogonal} to obtain the appropriate \mrep. This illustrates that \eqref{MOrthogonal} can be useful when given a $(\brep,{\sf O}) \in \mathfrak{B}_{\sf O}$. Given that simulating (pure-state) quantum trajectories requires less time and computer memory than a direct integration of the corresponding master equation \cite{WM10} one can expect applications of quantum trajectories to quantum information to continue in the foreseeable future.
%
%
%

%
%
%
%

\begin{acknowledgments}

This research was conducted by the \emph{Australian Research Council Centre of Excellence for Quantum Computation and Communication Technology} (project number CE110001029).

\end{acknowledgments}

\appendix

\section{VECTOR-OPERATOR ALGEBRA}
\label{VecOperatorAlg}

\subsection{Basic Operations}

In the same manner that we write a set of equations compactly using vectors and matrices in classical theories, we would also like to write a set of operator equations compactly. In order to do this we have to introduce a vector whose elements are operators which we call a vop, short for vector-operator. This appendix will make clear particular definitions that we adopt for vops and their algebra.

We begin with the definitions of the vop, its transpose, and its Hermitian conjugate. For an arbitrary vop $\hat{\bf A}$ consisting of $n$ components, 
\begin{align}
\label{VecOpDef1}
	\hat{\bf A} \equiv \left( \begin{array}{c} \hat{A}_{1} \\ \hat{A}_{2} \\ \vdots \\ \hat{A}_{n} \end{array} \right) 
	            \equiv \, \left( \hat{A}_{1}, \hat{A}_{2}, \ldots, \hat{A}_{n}\right)\tp \,, 
\end{align}
we define its Hermitian conjugate as
\begin{align}
\label{VecOpDef2}
	\hat{\bf A}\dg \equiv \,( \hat{A}\dg_{1}, \hat{A}\dg_{2}, \ldots, \hat{A}\dg_{n} ) \,. 
\end{align}
Note the Hermitian conjugate of $\hat{\bf A}$ is defined by transposing $\hat{\bf A}$ (as usually defined), and then taking the Hermitian conjugate of each element of $\hat{\bf A}$. We will say that a vop is Hermitian if and only if $\hat{\bf A}\tp = \hat{\bf A}\dg$. Matrix-operators 
also appear in this article and we will define the Hermitian conjugate of an $m \times n$ matrix-operator $\hat{\sf A}$ as follows:
\begin{align}
\label{MatrixOpDef}
	\hat{\sf A} \equiv {}&	\left( \begin{array}{cccc} \hat{\sf A}_{11}  &  \hat{\sf A}_{12}  &  \cdots   &  \hat{\sf A}_{1n}   \\
	                                                   \hat{\sf A}_{21}  &  \hat{\sf A}_{22}  &           &                     \\
	                                                        \vdots       &                    &  \ddots   &                     \\
	                                                   \hat{\sf A}_{m1}  &                    &           &  \hat{\sf A}_{mn}  
	                           \end{array} \right) \,, \\[0.20cm]
	\hat{\sf A}\dg \equiv {}& \left( \begin{array}{cccc} \hat{\sf A}\dg_{11}  &  \hat{\sf A}\dg_{21}  &  \cdots   &  \hat{\sf A}\dg_{m1}  \\
	                                                     \hat{\sf A}\dg_{12}  &  \hat{\sf A}\dg_{22}  &           &                        \\
	                                                           \vdots         &                       &  \ddots   &                        \\
	                                                     \hat{\sf A}\dg_{1n}  &                       &           &  \hat{\sf A}\dg_{mn}  
	                             \end{array} \right) \,.                                        
\end{align}
We will denote, for this appendix only, matrix-operators (which we refer to as mops), by San-serif letters with a hat on top. We will also refer to a $1 \times 1$ operator as a sop, short for scalar-operator. We retain the familiar properties of the Hermitian conjugate for complex matrices:
\begin{align}
\label{DoubleDaggerDefn}
	\big( \hat{\sf A}\tp \big)\dg = {}& \big( \hat{\sf A}\dg \big)\tp \,, \\
\label{MatrixOpHermitianConj}
	\big( \hat{\sf A} \hat{\sf B} \big)\dg = {}& \hat{\sf B}\dg \hat{\sf A}\dg \;,
\end{align}
where $\hat{\sf A}$ and $\hat{\sf B}$ are respectively $m \times n$ and $n \times k$ mops. Equation \eqref{DoubleDaggerDefn} crops up often so for convenience we define
\begin{align}
	\hat{\sf A}^{\ddag} = \big( \hat{\sf A}\tp \big)\dg \;.
\end{align}
It will also be useful for us to define the real and imaginary parts of an arbitrary sop $\hat{c}$ to be the Hermitian operators
\begin{align}
	\Re[\hat{c}] \equiv \frac{1}{2} \,(\hat{c} + \hat{c}\dg) \,,	\quad  \Im[\hat{c}] \equiv \frac{1}{2i} \,(\hat{c} - \hat{c}\dg) \,.
\end{align}
Since we also deal with complex matrices, we have also defined the real and imaginary parts of a complex matrix $C$ to be
\begin{align}
	\Re[C] \equiv \frac{1}{2}(C + C^{*}) \,, \quad  \Im[C] \equiv \frac{1}{2i}(C - C^{*}) \,,
\end{align}
where $C^*$ denotes the complex conjugate of $C$. We generalize the definitions of the real and imaginary parts of a sop to a mop:
\begin{align}
\label{ReMatrixOpDef}
	\Re\big[\hat{\sf A}\big] 
	\equiv {}& \left( \begin{array}{cccc} \Re[\hat{\sf A}_{11}]  &  \Re[\hat{\sf A}_{12}]  &  \cdots   &  \Re[\hat{\sf A}_{1n}]  \\
	                                      \Re[\hat{\sf A}_{21}]  &  \Re[\hat{\sf A}_{22}]  &           &                               \\
	                                            \vdots           &                         &  \ddots   &                               \\
	                                      \Re[\hat{\sf A}_{m1}]  &                         &           &  \Re[\hat{\sf A}_{mn}]  
	                 \end{array} \right) \\
	                           = {}& \frac{1}{2} \big( \,\hat{\sf A} + \hat{\sf A}^{\ddag} \,\big) \,, \\[0.15cm] 
\label{ImMatrixOpDef}
	\Im\big[\hat{\sf A}\big] 
	\equiv {}& \left( \begin{array}{cccc} \Im[\hat{\sf A}_{11}]  &  \Im[\hat{\sf A}_{12}]  &  \cdots   &  \Im[\hat{\sf A}_{1n}]  \\
	                                      \Im[\hat{\sf A}_{21}]  &  \Im[\hat{\sf A}_{22}]  &           &                               \\
	                                             \vdots          &                              &  \ddots   &                               \\
	                                      \Im[\hat{\sf A}_{m1}]  &                              &           &  \Im[\hat{\sf A}_{mn}]  
	                \end{array} \right) \\
                             = {}& \frac{1}{2i} \big( \,\hat{\sf A} - \hat{\sf A}^{\ddag} \,\big) \,.                                        
\end{align}
Note the necessity of taking both the transpose and Hermitian conjugate in \eqref{ReMatrixOpDef} and \eqref{ImMatrixOpDef} in order to retain an $m \times n$ mop.

Multiplication of $\hat{\sf A}$ by an arbitrary sop $\hat{c}$ is defined component-wise by
\begin{align}
\label{MultiplicationByScalar}
	\hat{c} \, \hat{\sf A} 
	\equiv {}&	\left( \begin{array}{cccc} \hat{c} \, \hat{\sf A}_{11}  &  \hat{c} \, \hat{\sf A}_{12}  &  \cdots   &  \hat{c} \, \hat{\sf A}_{1n}   \\
	                                       \hat{c} \, \hat{\sf A}_{21}  &  \hat{c} \, \hat{\sf A}_{22}  &           &                                 \\
	                                                   \vdots           &                               &  \ddots   &                                 \\
	                                       \hat{c} \, \hat{\sf A}_{m1}  &                               &           &  \hat{c} \, \hat{\sf A}_{mn}  
	                     \end{array} \right) \,,
\end{align}
and similarly for the action of an arbitrary superoperator ${\cal N}$, 
\begin{align}
\label{SuperopMultiplication}
	{\cal N} \, \hat{\sf A} 
	\equiv {}&	
	\left( \begin{array}{cccc} {\cal N} \, \hat{\sf A}_{11}  &  {\cal N} \, \hat{\sf A}_{12}  &  \cdots   &  {\cal N} \, \hat{\sf A}_{1n}   \\
	                           {\cal N} \, \hat{\sf A}_{21}  &  {\cal N} \, \hat{\sf A}_{22}  &           &                                  \\
	                                         \vdots          &                                &  \ddots   &                                  \\
	                           {\cal N} \, \hat{\sf A}_{m1}  &                                &           &  {\cal N} \, \hat{\sf A}_{mn}  
	                     \end{array} \right)  \;.
\end{align}
It is straightforward to see (and will be useful to remember), from \eqref{MultiplicationByScalar} that
\begin{align}
	\hat{c} \, \hat{\sf A}\tp = {}& \big( \hat{c} \, \hat{\sf A} \big)^{\!\top}  \;, \\
	\big( C \hat{\sf A} \big)\tp = {}& \hat{\sf A}\tp C\tp \;,
\end{align}
for any $k \times m$ matrix $C$.

Two special cases of \eqref{MatrixOpHermitianConj} that we are interested in are the inner- and outer-products between vops for which \eqref{MatrixOpHermitianConj} reads 
\begin{align}
\label{HermitianConjOfInnerProd}
	\big( \hat{\bf A}\!\tp \hat{\bf B} \big)\dg = \hat{\bf B}\dg \hat{\bf A}^{\ddag} \,,  \\
\label{HermitianConjOfOuterProd}
	\big( \hat{\bf A} \hat{\bf B}\tp \big)\dg = \hat{\bf B}^{\ddag} \hat{\bf A}\dg \,.   
\end{align}
The following identities will also be useful for manipulating vop products
\begin{align}
	         \big( \vop{A}\tp \vop{B} \big) \vop{C} = {}& \big( \vop{A}\tp \vop{B} \vop{C}\tp \big)\tp  \;, \\
	\big( \vop{A} \vop{B}\tp \big)^{\!\top} \vop{C} = {}& \Big[ \vop{A}^{\!\top} \big( \vop{B} \vop{C}\tp \big)^{\!\top} \;\!\Big]\tp  \;.
\end{align}

The tensor product for vops are defined analogous to standard multiplication rules. If $\hat{\bf A}$ and $\hat{\bf B}$ have different dimensions then we define
\begin{align}
	\hat{\bf A} \otimes \hat{\bf B}\tp 
	\equiv \left( \begin{array}{cccc} \hat{A}_1 \otimes \hat{B}_1  &  \hat{A}_1 \otimes \hat{B}_2  &  \cdots   &  \hat{A}_1 \otimes \hat{B}_n  \\
	                                  \hat{A}_2 \otimes \hat{B}_1  &  \hat{A}_2 \otimes \hat{B}_2  &           &                               \\
	                                               \vdots          &                               &  \ddots   &                               \\
	                                  \hat{A}_m \otimes \hat{B}_1  &                               &           &  \hat{A}_m \otimes \hat{B}_n  
	                  \end{array} \right) \;.
\end{align}
When $\vop{A}$ and $\vop{B}$ have the same dimension then 
\begin{align}
	\hat{\bf A}^{\!\top} \otimes \hat{\bf B} \equiv \sum_{k} \hat{A}_k \otimes \hat{B}_k \;.
\end{align}

We will assign the following symbols for identity operations with respect to multiplication: 
\begin{itemize}

\item
$\hat{1}$ for a sop identity, 

\item
$\ID{N}$ for an $N \times N$ identity matrix, 

\item
$\mopID{N}$ for a mop identity (defined by $\mopID{N}=\ID{N}\hat{1}$), 


\end{itemize}
While it is common to find special symbols for identity operations for multiplication in the literature, it is also of common practice to \emph{not} denote the identity with respect to addition by any distinguished symbol except for 0. We will follow this convention as well.

\subsection{Commutativity}

\subsubsection{Sop-brackets}

Unlike normal vectors the order of the vops in an inner-product may not be changed in general, 
\begin{align}
	\vop{B}\tp \vop{A} \ne \vop{A}\!\tp \vop{B} \;,
\end{align}
and this motivates us to define the sop-bracket
\begin{align}
\label{ScalarBracket}
	\big\lceil \vop{A} , \vop{B} \big\rfloor = \vop{A}\tp \vop{B} - \vop{B}\tp \vop{A} \;,
\end{align}
so called because it maps two vops to a sop. We then have the following sufficient (but not necessary) condition for reordering the inner-product,
\beq
\label{ScalarSuffCond}
	\forall \;k , \; [\hat{A}_{k},\hat{B}_{k}] = 0  \implies  \big\lceil \vop{A} , \vop{B} \big\rfloor = 0 \,.	
\eeq
That $[\hat{A}_{k},\hat{B}_{k}] = 0$ is not necessary for $\big\lceil \vop{A},\vop{B} \big\rfloor = 0$ can be seen in the following example, 
\beq
	\hat{\bf A}=(\hat{q},\hat{p})\tp \,, \quad \hat{\bf B}=(\hat{p},\hat{q})\tp \,, \quad [\hat{q},\hat{p}]=i\hbar \,.
\eeq
Here $\hat{A}_{1}$ does not commute with $\hat{B}_{1}$, and $\hat{A}_{2}$ does not commute with $\hat{B}_{2}$ but we still have $\hat{\bf A}\!\tp \hat{\bf B} = \hat{\bf B}\tp \hat{\bf A}$. Note that we have not overloaded the usual notation for commutator brackets in \eqref{ScalarBracket} because while the sop-bracket satisfies the following (which is easy to check)
\begin{align}
	\big\lceil \vop{A} , \vop{B} \big\rfloor = {}& - \big\lceil \vop{B} , \vop{A} \big\rfloor  \;,  \\
	\big\lceil \vop{A} + \vop{B} , \vop{C} \big\rfloor = {}& \big\lceil \vop{A} , \vop{C} \big\rfloor + \big\lceil \vop{B} , \vop{C} \big\rfloor \;,
\end{align}
it cannot satisfy a property analogous to the Jacobi identity which the standard commutator does.

The inner-product is a sop so by definition $\vop{A}\tp \vop{B} = ( \vop{A}\!\tp \vop{B})\!\tp$. Thus \eqref{ScalarSuffCond} may also be seen as a sufficient condition for $(\vop{A}\!\tp \vop{B})\!\tp = \vop{B}\tp \vop{A}$.

\subsubsection{Mop-brackets}

For the vop outer-product we do have a necessary and sufficient condition for writing the transpose of $\hat{\bf A} \hat{\bf B}\tp$ as a product of transposes:
\bqa
\label{VecOpCommutator}
	\forall \, j,k , \; [\,\hat{A}_{j},\hat{B}_{k}\,] = 0   
	\; \Longleftrightarrow \; \big( \hat{\bf A} \, \hat{\bf B}\tp \big)\tp = \hat{\bf B} \, \hat{\bf A}\tp \,.
\eqa
Therefore we define the mop-bracket 
\begin{align}
\label{MatrixBracket}
	\big\lfloor \hat{\bf A}, \hat{\bf B} \big\rceil \equiv  \hat{\bf A} \hat{\bf B}\tp - \big( \hat{\bf B} \hat{\bf A}\tp \big)\tp  \;,
\end{align}
which maps vops to mops, and refer to $\hat{\bf A}$ and $\hat{\bf B}$ as commuting vops if and only if $\big\lfloor \hat{\bf A}, \hat{\bf B} \big\rceil = 0$. It is easy to see that this satisfies
\begin{align}
	\big\lfloor \hat{\bf A}, \hat{\bf B} \big\rceil = {}& - \big\lfloor \hat{\bf B}, \hat{\bf A} \big\rceil\tp  \;, \\
	\big\lfloor \hat{\bf A} + \hat{\bf B}, \hat{\bf C} \;\! \big\rceil = {}& \big\lfloor \hat{\bf A}, \hat{\bf C} \;\! \big\rceil 
	                                                                         + \big\lfloor \hat{\bf B}, \hat{\bf C} \;\! \big\rceil \;.
\end{align}
The symbol for the mop-bracket in \eqref{MatrixBracket} has the ceiling ($\rceil$) on $\vop{B}$ to remind us of $\vop{A} \vop{B}\tp$, which is a mop. For the sop-bracket of \eqref{ScalarBracket} the ceiling is placed on $\vop{A}$ to remind us of $\vop{A}\!\tp \vop{B}$. Note the commutativity of vops thus defined means that an arbitrary vop will not necessarily commute with itself as that is a question about its components.

From the above one may verify the following formulae which help simplify expressions containing mop-brackets: For any matrix $C$ (with the appropriate dimensions)
\begin{align}
	\big\lfloor C \vop{A}, \vop{B} \big\rceil = {}& C \big\lfloor \vop{A}, \vop{B} \big\rceil \;,  \\
	\big\lfloor \vop{A}, C \vop{B} \big\rceil = {}& \big\lfloor \vop{A}, \vop{B} \big\rceil C\tp \;. 
\end{align}
For any sop $\hat{c}$,
\begin{align}
\label{ScalarVop1}
	\big\lfloor \hat{c},  \vop{A} \big\rceil = {}& \big[ \hat{c}, \vop{A}^{\!\top} \big] \;,  \\
\label{ScalarVop2}
	\big\lfloor \vop{A}, \hat{c} \;\!\big\rceil = {}& \big[ \vop{A}, \hat{c} \;\!\big] \;.
\end{align}
Note that it is the standard commutator which appears on the RHS of \eqref{ScalarVop1} and \eqref{ScalarVop2}. We also have
\begin{align}
	\big\lfloor \vop{A} \;\! \hat{c}, \vop{B} \big\rceil = {}& \vop{A} \big[ \hat{c},\vop{B}\tp \big] 
	                                                           + \big\lfloor \vop{A}, \vop{B} \big\rceil \hat{c} \;,	\\
	\big\lfloor \hat{c} \;\! \vop{A}, \vop{B} \big\rceil = {}& \hat{c} \big\lfloor \vop{A}, \vop{B} \big\rceil 
	                                                           + \Big( \big[ \hat{c},\vop{B} \big] \vop{A}\tp \Big)^{\!\top} \;,	\\
	\big\lfloor \vop{A}, \vop{B} \;\! \hat{c} \;\!\big\rceil = {}& \Big( \vop{B} \big[ \vop{A}^{\!\top},\hat{c} \big] \Big)^{\!\top} 
	                                                               + \big\lfloor \vop{A}, \vop{B} \big\rceil \hat{c} \;,  \\
	\big\lfloor \vop{A}, \hat{c} \;\! \vop{B} \big\rceil = {}& \hat{c} \big\lfloor \vop{A}, \vop{B} \rceil 
	                                                           + \big[ \vop{A}, \hat{c} \;\! \big] \vop{B}\tp  \;.
\end{align}

\subsection{Trace and Average}

Since both matrices and operators appear we will denote an operator-trace by ${\rm Tr}[\hat{c}]$ and the trace of a matrix $C$ by ${\rm tr}[C]$. This distinction is especially important for operators with a continuous eigenvalue spectrum where the operator-matrix analogy breaks down. The trace of an $m \times n$ mop is a matrix in $\mathbb{C}^{m \times n}$
\begin{align}
\label{MatrixOpTrace}
	{\rm Tr}\big[ \hat{\sf A} \big]  
	  \equiv	\left( \begin{array}{cccc} {\rm Tr}[\hat{\sf A}_{11}] & {\rm Tr}[\hat{\sf A}_{12}] & \cdots & {\rm Tr}[\hat{\sf A}_{1n}]   \\
	                                     {\rm Tr}[\hat{\sf A}_{21}] & {\rm Tr}[\hat{\sf A}_{22}] &        &                              \\
	                                              \vdots            &                            & \ddots &                              \\
	                                     {\rm Tr}[\hat{\sf A}_{m1}] &                            &        & {\rm Tr}[\hat{\sf A}_{mn}]                                    \end{array} \right) \,.
\end{align}
From this it should be simple to see that
\begin{align}
\label{MatrixOpTrace1}
	{\rm Tr}\big[ \hat{\sf A}\tp \big] = {}& \big( {\rm Tr}\big[ \hat{\sf A} \big] \big)^{\!\top} \,, \\
\label{MatrixOpTrace2}
	\big( {\rm Tr}[ \hat{\sf A} ] \big)^{\!*} = {}& {\rm Tr}\big[ \hat{\sf A}^{\ddag} \;\!\big] \,.
\end{align}
Particular cases of \eqref{MatrixOpTrace1} and \eqref{MatrixOpTrace2} which we are interested in are
\begin{align}
\label{HermitianConjOfMatrixOpTrace}
	\big(\;\! {\rm Tr}[ \hat{\bf A} ] \,\big)\dg = {}& {\rm Tr}[ \hat{\bf A}^{\!\dag} ] \;, \\
	\big( {\rm Tr}\big[ \hat{\bf A} \hat{\bf B}\tp \big] \big)^{*} = {}& {\rm Tr}\big[ \hat{\bf A}^{\ddag} \hat{\bf B}\dg \big]	\,.
\end{align}
If $\hat{\sf A}$ is any $m \times n$ mop and $\hat{\sf B}$ is $n \times k$, then we have the ``cyclic'' property
\begin{align}
\label{MatrixOpTrace3}
	{\rm Tr}\big[ ( \hat{\sf A} \hat{\sf B})\!\tp \big] = {\rm Tr}\big[ \hat{\sf B}\tp \hat{\sf A}\tp \big] \;.
\end{align}
Special cases of the ``cyclic'' property \eqref{MatrixOpTrace3} are
\bqa
	{\rm Tr}\big[\;\! \hat{c} \;\! \hat{\bf A} \big] &=&  {\rm Tr}\big[ \hat{\bf A} \;\! \hat{c} \;\! \big] \,, \\
\label{VecOpOuterProdTrace}
	{\rm Tr}\big[ \big( \hat{\bf A} \hat{\bf B}\tp \big)^{\!\top} \big] &=& {\rm Tr}\big[ \hat{\bf B} \hat{\bf A}\!\tp \big] \,.
\eqa

On some occasions we will want to permute a product of three vops inside a trace and the following identity may be verified,
\begin{align}
	{\rm Tr}\big[ \vop{A} \;\! \vop{B}\tp \vop{C} \;\!\big]
	=  {}& {\rm Tr}\big[ \vop{B}\tp \vop{C} \;\! \vop{A} \big]
	= \Tr \big[ (\vop{C} \vop{A}\tp)\tp \vop{B} \big] \;.
%
%
\end{align}

An important tool from quantum statistics that we will use is the quantum regression formula, also known as the quantum regression theorem, formulated by Lax \cite{Lax63, Lax67}. A good discussion of this can be found in \cite{Car02}. In essence this result expresses the average of a product of two operators, each at a different time, as a trace in the \sch\ picture. To write down the vop counterpart of this result we first recall the regression formulae for sops: Given any two operators $\hat{A}$ and $\hat{B}$ in the system Hilbert space, and the solution to the Markovian master equation $\dot{\rho}={\cal L}\rho$, 
\begin{align}
\label{OpRegress1}
	\big\langle \hat{A}(t) \,\hat{B}(t+\tau) \big\rangle 
	 & = {\rm Tr} \!\left\{ \hat{B}(0) \, e^{{\cal L}\tau} \! \left[ \rho(t) \,\hat{A}(0) \right] \right\} \,, \\
\label{OpRegress2}
	\big\langle \hat{A}(t+\tau) \,\hat{B}(t) \big\rangle 
	 & = {\rm Tr} \!\left\{ \hat{A}(0) \, e^{{\cal L}\tau} \! \left[ \hat{B}(0) \,\rho(t) \right] \right\} \,.
\end{align}
for any $\tau > 0\,$. Let us define the quantum average of a mop
$\hat{\sf A}$ at time $t$ as 
\begin{align}
	\langle \hat{\sf A} \rangle = {\rm Tr}[ \rho(t) \hat{\sf A}] \,.
\end{align}
A corollary which follows from this definition and \eqref{HermitianConjOfMatrixOpTrace} is 
\beq
	\big\langle \hat{c} \, \hat{\bf A} \big\rangle\dg = \big\langle \hat{\bf A}^{\!\dag} \, \hat{c}\dg \big\rangle \,,
\eeq
obtained by letting $\hat{\bf A} \rightarrow \rho \hat{c} \hat{\bf A}$ in \eqref{HermitianConjOfMatrixOpTrace}. We can now extend \eqref{OpRegress1} and \eqref{OpRegress2} to
\begin{align}
\label{VecOpRegress1}
	\big\langle \hat{\bf A}(t) \,\hat{\bf B}\tp(t+\tau) \big\rangle 
	 & = \left( {\rm Tr} \!\left\{ \hat{\bf B}(0) \, e^{{\cal L}\tau} \! \left[ \rho(t) \,\hat{\bf A}\tp(0) \right] \right\} \right)\tp , \\
\label{VecOpRegress2}
	\big\langle \hat{\bf A}(t+\tau) \,\hat{\bf B}\tp(t) \big\rangle 
	 & = {\rm Tr} \!\left\{ \hat{\bf A}(0) \, e^{{\cal L}\tau} \! \left[ \hat{\bf B}\tp(0) \,\rho(t) \right] \right\} \,.
\end{align}

\subsection{Superoperators of Special Interest}

When considering continuous measurements terms of certain forms arise frequently. Here we will define superoperators whose forms allow us to write these frequently occuring terms compactly. For an arbitrary sop $\hat{s}$ we define
\begin{align}
\label{SuperOpJ1}
	{\cal J}[\hat{A}] \hat{B} \equiv \hat{A} \;\! \hat{B} \;\! \hat{A}\dg \;.
\end{align}
Note that $\hat{A}$ here should be treated as a parameter and $\hat{B}$ the input for ${\cal J}[\hat{A}]$, the output of the superoperator is another sop given by the RHS of \eqref{SuperOpJ1}. The definition \eqref{SuperOpJ1} thus implies that ${\cal J}[\hat{A}\dg] \hat{B} = \hat{A}\dg \;\! \hat{B} \;\! \hat{A}$.

Often we encounter a sum of terms in the form of \eqref{SuperOpJ1} so we define
\begin{align}
\label{SuperOpJ2}
	{\cal J}[\vop{A}] \hat{B} \equiv \sum_k {\cal J}[\hat{A}_k] \hat{B} = \sum_k \, \hat{A}_k \;\! \hat{B} \, \hat{A}\dg_k 
	                                                                    = \vop{A}\tp \hat{B} \, \vop{A}\ddg  \;.
\end{align}
From \eqref{SuperopMultiplication} and \eqref{SuperOpJ2} the action of ${\cal J}[\vop{A}]$ on a vop produces another vop, given by
\begin{align}
\label{SuperOpJ3}
	{\cal J}[\vop{A}] \;\! \vop{B} = \sum_k  \hat{A}_k \;\! \vop{B} \, \hat{A}\dg_k = \big( \vop{A} \;\! \vop{B}\tp \big)^{\!\top} \vop{A}\ddg \;.
\end{align}
This also gives	${\cal J}[\vop{A}\ddg] \vop{B} = \big( \vop{A}\ddg \vop{B}\tp \big)^{\!\top} \vop{A}\,$. Here we note that ${\cal J}[\vop{A}\ddg]$ is formally equivalent to its superoperator adjoint \footnote{The standard definition of the adjoint of a superoperator ${\cal N}$ \cite{BP02}, is another superoperator ${\cal N}\dg$, such that for any sop $\hat{A}$, and any state $\rho$, $\Tr \big[ \hat{A} ({\cal N} \rho) \big] = \Tr \big[ ({\cal N}\dg \hat{A}) \rho \big]$. We can generalize $\hat{A}$ in this definition to a vop $\vop{A}$ so that ${\cal N}\dg$ is such that $\Tr \big[ \vop{A} ({\cal N} \rho) \big] = \Tr \big[ ({\cal N}\dg \vop{A}) \rho \big]$. From this it follows that $\big({\cal J}[\vop{A}]\big)\dg = {\cal J}[\vop{A}\ddg]$.}.

From \eqref{SuperOpJ1} we can define what is sometimes referred to as the ``dissipator'' \cite{BP02}
\begin{align}
\label{SupeOpD1}
	{\cal D}[\hat{A}] \hat{B} \equiv {\cal J}[\hat{A}] \hat{B} - \frac{1}{2} \, \big\{ \hat{A}\dg \hat{A}, \hat{B} \big\}  \;,
\end{align}
where $\big\{\hat{A},\hat{B}\big\} \equiv \hat{A} \hat{B} + \hat{B} \hat{A}$. When $\hat{A}=\hat{A}\dg$ one may prefer to write the dissipator as a nested commutator,
\begin{align}
\label{NestedCommutatorForm}
	{\cal D}[\hat{A}] \hat{B} = - \frac{1}{2} \, \big[ \hat{A}, [ \hat{A},\hat{B} ] \big] \;.
\end{align}
As in \eqref{SuperOpJ2}, we shorthand a sum of dissipators by 
\begin{align}
\label{SuperOpD2}
	{\cal D}[\vop{A}] \hat{B} \equiv \sum_k {\cal D}[\hat{A}_k] \hat{B}  \;.
\end{align}
Equations \eqref{SuperopMultiplication} and \eqref{SuperOpD2} then permit us to write
\begin{align}
\label{SuperOpD3}
	{\cal D}[\vop{A}] \;\! \vop{B} = {}& {\cal J}[\vop{A}] \vop{B} - \frac{1}{2} \, \big\{ \vop{A}\dg \vop{A}, \vop{B} \big\}  \nn \\
	                          = {}& \big( \vop{A} \;\! \vop{B}\tp \big)^{\!\top} \vop{A}\ddg - \frac{1}{2} \, \big\{ \vop{A}\dg \vop{A},\vop{B} \big\} \;.
\end{align}
In the case when $\vop{A}=\vop{A}\ddg$ the vop-generalization of \eqref{NestedCommutatorForm} is 
\begin{align}
	{\cal D}[\vop{A}] \;\! \vop{B} 
	= - \frac{1}{2} \, \Big\{ \big( \vop{A}\tp \mbrac{\vop{A}}{\vop{B}} \big)\tp + \mbrac{\vop{B}}{\vop{A}} \vop{A} \Big\} \;.
\end{align}

The appearance of a dissipator ${\cal D}[\hat{A}]$ in the master equation is associated with the coupling of the system to the environment via $\hat{A}$. If the system is also under continuous observation then measurement back-action terms arise and they can be written concisely by defining
\begin{align}
\label{SuperOpH1}
	{\cal H}[\hat{A}] \hat{B} \equiv \hat{A} \hat{B} + \hat{B} \hat{A}\dg - \Tr \big( \hat{A} \hat{B} + \hat{B} \hat{A}\dg \big) \hat{B} \;.
\end{align}
We will invariably be considering ${\cal H}[{\bf C}\tp \vop{A}]$ with ${\bf C}={\bf C}^*$. In this case one may prefer to generalize \eqref{SuperOpH1} to
\begin{align}
\label{SuperOpH2}
	{\cal H}[\vop{A}] \hat{B} \equiv \vop{A} \hat{B} + \hat{B} \vop{A}\ddg - \Tr \big[ \vop{A} \hat{B} + \hat{B} \vop{A}\ddg \big] \hat{B}
\end{align}
and write 
\begin{align}
	{\cal H}[{\bf C}\tp \vop{A}] = {\bf C}\tp {\cal H}[\vop{A}]  \;.
\end{align}
Note that \eqref{SuperOpH1} and \eqref{SuperOpH2} each contain a term which is nonlinear in $\hat{B}$. One may find it useful to also define the linear versions of \eqref{SuperOpH1} and \eqref{SuperOpH2},
\begin{align}
	\bar{\cal H}[\hat{A}] \hat{B} \equiv {}& \hat{A} \hat{B} + \hat{B} \hat{A}\dg \;, \\
	\bar{\cal H}[\vop{A}] \hat{B} \equiv {}& \vop{A} \hat{B} + \hat{B} \vop{A}\ddg  \;.
\end{align}

\section{DERIVATION OF \erf{PurityInTermsOfAlpha}}
\label{AppendixForM}

We first write (suppressing the time-dependence)
\begin{align}
\label{PurityWithPlancksConstant}
	\Tr\big[ (\rho + d\rho_{\rm c})^2 \big] = 1 + \frac{2}{\hbar} \, \Tr\big[ \rho \, (\hbar \;\! d\rhoc) \big] 
	                                          + \frac{1}{\hbar^2} \, \Tr\big[ (\hbar \;\! d\rhoc)^2 \big] \;,
\end{align}
where the change in the state is given by
\begin{align}
\label{MrepSME}
	\hbar \, d\rhoc = {\cal L} \;\! \rho \, dt + {\cal H}[d{\bf w}\tp {\sf M}\dg \hc] \rho  \;.
\end{align}
Recall that ${\cal L}$ is defined by \eqref{LindbladForm}. We will assume that $\rho(t)$ is unconditioned and such that $\Tr[\rho(t)]=1$ and $\rho^2(t)=\rho(t)$.

We first work out the second term in \eqref{PurityWithPlancksConstant}. Taking the trace of \eqref{MrepSME} against $\rho$,
\begin{align}
\label{2ndTermChangeInPurity1}
	\Tr\big\{ \rho \;\!(\hbar \, d\rhoc) \big\} = \Tr\big\{ \rho \;\! {\cal L} \;\! \rho \big\} dt 
	                                             + \Tr\big\{ \rho {\cal H}[d{\bf w}\tp {\sf M}\dg \hc] \rho \big\} \;.
\end{align}
From \eqref{SuperOpH1} we find, for any $\hat{A}$,
\begin{align}
	\Tr\big\{ \rho \, {\cal H}[\hat{A}] \rho \big\} 
%
             = 0 \;.
\end{align}
From \eqref{LindbladForm} we get
\begin{align}
	\Tr\big\{ \rho {\cal L} \rho \big\} 
%
	                                    = \ban{\hc\dg} \ban{\hc} - \ban{\hc\dg \hc}  \;.
\end{align}
Therefore \eqref{2ndTermChangeInPurity1} is
\begin{align}
\label{2ndTermChangeInPurity2}
	\Tr\big\{ \rho \;\!(\hbar \, d\rhoc) \big\} = \Tr\big\{ \rho \;\! {\cal L} \rho \big\} dt 
	                                            = \big( \ban{\hc\dg} \ban{\hc} - \ban{\hc\dg \hc} \big) dt  \;.
\end{align}

To order $dt$ the third term in \eqref{PurityWithPlancksConstant} is proportional to  
\begin{align}
	\bTr{(\hbar \;\! d\rhoc)^2\;\!} = \bTr{ \big( {\cal H}[d{\bf w}\tp {\sf M}\dg \hc] \rho \big)^2\;\!}  \;.
\end{align}
For any $\hat{A}$,
\begin{align}
	{}& \hspace{-0.5cm}  \big( {\cal H}[\hat{A}] \rho \big)^2  \nn \\
%
   = {}& (\hat{A} \rho) (\hat{A} \rho) + (\hat{A} \rho) (\rho \hat{A}\dg) 
         - (\hat{A} \rho) \;\! \bTr{\hat{A} \rho + \rho \hat{A}\dg} \rho  \nn \\
       & + (\rho \hat{A}\dg) (\hat{A} \rho) + (\rho \hat{A}\dg) (\rho \hat{A}\dg) 
         - (\rho \hat{A}\dg) \;\! \bTr{\hat{A} \rho + \rho \hat{A}\dg} \rho \nn \\
       & + \bTr{\hat{A} \rho + \rho \hat{A}\dg} \bTr{\hat{A} \rho + \rho \hat{A}\dg} \rho^2  \;.
\end{align}
Taking the trace and using $\rho^2=\rho$,
\begin{align}
	\Tr\big\{ \big( {\cal H}[\hat{A}]\rho \big)^2 \big\} 
%
	= 2 \big( \ban{\hat{A}\dg \hat{A}} - \ban{\hat{A}} \ban{\hat{A}} \big)  \;.                    
\end{align}
This gives, for $\hat{A}=d{\bf w}\tp {\sf M}\dg \hc$,
\begin{align}
\label{3rdTermChangeInPurity}
	\Tr\big\{ \big( {\cal H}[d{\bf w}\tp {\sf M}\dg \hc\;\!] \;\!\rho \big)^2 \big\} 
	= 2 \;\! \big( \ban{\hc\dg \MMdg \hc} - \ban{\hc} \MMdg \ban{\hc} \big) dt  \,.
\end{align}
Thus substituting \eqref{2ndTermChangeInPurity2} and \eqref{3rdTermChangeInPurity} into \eqref{PurityWithPlancksConstant},
\begin{align}
	{}& \hspace{-0.5cm} \bTr{(\rho+d\rhoc)^2\;\!}  \nn \\
%
%
	= {}& 1 + \frac{2}{\hbar} \, 
	      \tr\big[ \ban{\big(\hc-\an{\hc}\!\big)\ddg \big(\hc-\ban{\hc}\big)^{\!\top}} \big( \ID{L} - \MMdg/\hbar \big)^{\!\top} \big] dt  \nn \\
	= {}& 1 + \tr\big[ H \big( \ID{L} - \MMdg/\hbar \big)^{\!\top} \big] \, dt \;.
\end{align} 
The matrix $H$ was defined in \eqref{CovarianceOfLindbladOperators}.

\section{\erf{MOrthogonal} FOR $L=1$}
\label{FactorizationExample}

\begin{figure*}
\includegraphics[width=17cm]{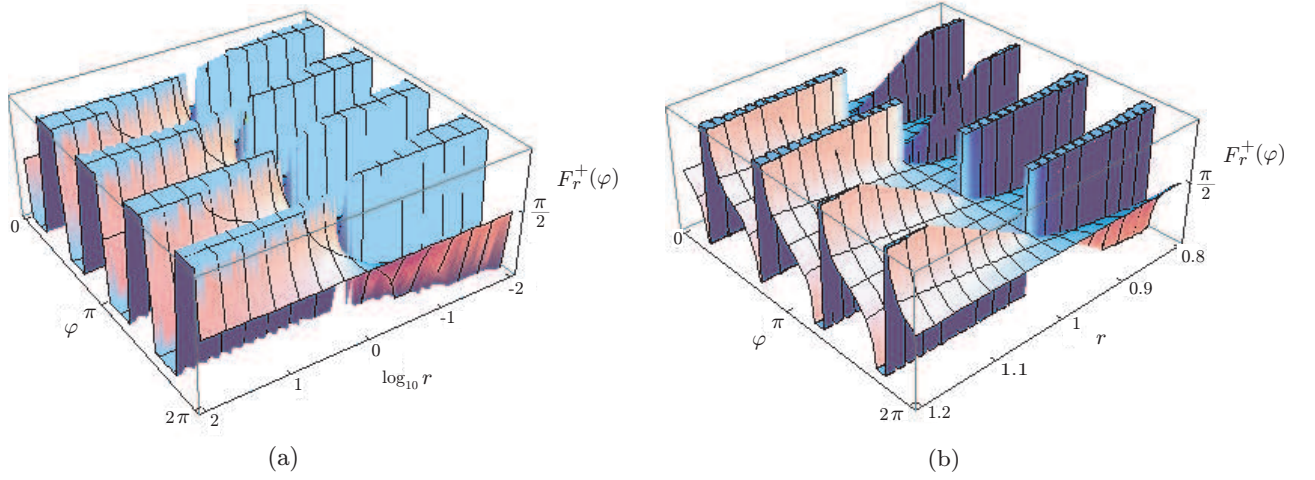} 
\caption{$F^{+}_{r}(\varphi)$ for different values of $r\equiv|m_1/m_2|^2$ and $\varphi$. (a) Behaviour of $F^{+}_r(\varphi)$ for large values of $r$. (b) $F^{+}_r(\varphi)$ around $r=1$. $F^{+}_r(\varphi)$ becomes $\pi/2$ at $r=1$ (see \eqref{F1}) and has a ``twist'' in crossing $r=1$. Remember that $|m_1|^2$ and $|m_2|^2$ are constrained by \eqref{Mconstraint} so given an $|m_2|^2 \ne 0$, $|m_1|^2$ cannot be an arbitrary multiple of $|m_2|^2$ with $|m_2|^2$ staying constant. The limit $r \to \infty$ is thus equivalent to the limits $|m_1|^2 \to 1$ and $|m_2|^2 \to 0$. Similary, if we are given an $|m_1|^2 \ne 0$, then $|m_2|^2$ cannot be an arbitrary multiple of $|m_1|^2$ with $|m_1|^2$ staying constant so the limit $r \to 0$ is equivalent to the limits $|m_1|^2 \to 0$ and $|m_2|^2 \to 1$.}
\label{FactorizationA}
\end{figure*}
We can show that for $L=1$ it is always possible to find a decomposition in the form of \eqref{MOrthogonal} for an arbitrary ${\sf M}$. The most general $2 \times 2$ orthogonal matrix can be parameterized by a single variable $\varphi$,
\begin{align}
\label{2X2O}
	{\sf O} = \tbt{\cos \varphi}{\sin \varphi}{\mp \sin \varphi}{\pm \cos \varphi}
\end{align}
where the sign in the second row corresponds to $\det({\sf O})=\pm 1$. This gives, from \eqref{MOrthogonal},
\begin{align}
\label{MgivenBforL=1}
	{\sf M}\tp = {}& \rt{\hbar\;\!\eta} e^{i\phi} \tbo{\rt{\theta} \cos \varphi \pm i \rt{\bar{\theta}} \sin \varphi}
	                                              {\rt{\theta} \sin \varphi \mp i \rt{\bar{\theta}} \cos \varphi}  \;.
%
\end{align}
%
%
%
%
If we define
\begin{align}
	{\sf M}\tp = \tbo{m_1}{m_2} = \tbo{|m_1|\;\!e^{i\alpha_1}}{|m_2|\;\!e^{i\alpha_2}} \;,
\end{align}
then our problem is to find $(\eta,\phi,\theta,\varphi)$ such that 
\begin{align}
\label{Equivalence1}
	|m_1| \;\! e^{i\alpha_1} = {}& \rt{\hbar\;\!\eta} e^{i\phi} \big( \rt{\theta} \cos \varphi \pm i \rt{\bar{\theta}} \sin \varphi \big)  \\
\label{Equivalence2}	
	|m_2| \;\! e^{i\alpha_2} = {}& \rt{\hbar\;\!\eta} e^{i\phi} \big( \rt{\theta} \sin \varphi \mp i \rt{\bar{\theta}} \cos \varphi \big)  \;,
\end{align}
given $m_1$, $m_2$, which determine $\eta$ by
\begin{align}
\label{Mconstraint}
 |m_1|^2 +|m_2|^2 = \hbar \, \eta  \;.
\end{align}
Equations \eqref{Equivalence1} and \eqref{Equivalence2} are true if and only if the modulus and argument (phase) of each side are equal, i.e.
\begin{align}
\label{EqnForm1}
	|m_1|^2 = \hbar \;\! \eta \;\! [ \theta \cos^2 \varphi + \bar{\theta} \sin^2 \varphi ] \;, \\
\label{EqnForm2}
	|m_2|^2 = \hbar \;\! \eta \;\! [ \theta \sin^2 \varphi + \bar{\theta} \cos^2 \varphi ] \;,
\end{align}
and 
\begin{align}
\label{EqnForalpha1}
	\alpha_1 = {}& \phi + \arg\big( \rt{\theta} \cos \varphi \pm i \rt{\bar{\theta}} \sin \varphi \big) \;, \\
\label{EqnForalpha2}
	\alpha_2 = {}& \phi + \arg\big( \rt{\theta} \sin \varphi \mp i \rt{\bar{\theta}} \cos \varphi \big)  \;.
\end{align}
Rearranging \eqref{EqnForm1} we find
\begin{align}
\label{ThetaWithEfficiency}
	\theta = \frac{|m_1|^2 - \hbar \;\! \eta \sin^2 \varphi}{\hbar \;\! \eta \cos(2\;\!\varphi)} \;.
\end{align}
If we now substitute \eqref{Mconstraint} into \eqref{ThetaWithEfficiency} and define $r = (|m_1|/|m_2|)^2$, then we get
\begin{align}
\label{ThetaSoln}
	\theta_r(\varphi) \equiv \frac{r \cos^2 \varphi - \sin^2 \varphi}{(r + 1) (\cos^2 \varphi - \sin^2 \varphi)}  \;.
\end{align}
Note that $\varphi$ must also ensure $0\le\theta\le1$. For small $r$ we find, from \eqref{ThetaSoln},
\begin{align}
	\theta_0(\varphi) =  \frac{-\sin^2 \varphi}{\cos^2 \varphi - \sin^2 \varphi} 
\end{align}
while for $r$ large $\theta_r(\varphi)$ approaches
\begin{align}
	\theta_\infty(\varphi) \equiv \frac{\cos^2 \varphi}{\cos^2 \varphi - \sin^2 \varphi} \;.
\end{align}
At these extremes the only values of $\varphi$ that enforce $\theta \in [0,1]$ are integer multiples of $\pi/2$, for which $\theta$ is either $0$ or $1$.

If we now attempt to eliminate $\theta$ in \eqref{EqnForm2} by substituting in \eqref{ThetaSoln} we simply arrive at \eqref{Mconstraint}, which is independent of $\varphi$. Thus \eqref{EqnForm1} and \eqref{EqnForm2} are solved simultaneously with any $\varphi$ and \eqref{ThetaSoln}. The value of $\varphi$ is determined from solving \eqref{EqnForalpha1} and \eqref{EqnForalpha2} simultaneously. First eliminating $\phi$ we see that $\varphi$ must satisfy
\begin{align}
\label{VarphiDefn}
	\delta \equiv \alpha_1 - \alpha_2 =  F^{\pm}_{r}(\varphi) \;,
\end{align}
where we have defined the argument of $m_1/m_2$ as
\begin{align}
\label{Fpm}
	F^{\pm}_{r}(\varphi) \equiv {}& \arg\left[ \frac{\rt{\theta_{r}(\varphi)} \cos \varphi \pm i \rt{1-\theta_{r}(\varphi)} \sin \varphi}
	                                                    {\rt{\theta_{r}(\varphi)} \sin \varphi \mp i \rt{1-\theta_{r}(\varphi)} \cos \varphi} \right] \;.
\end{align}
We have written $\theta_{r}(\varphi)$ in \eqref{Fpm} to signify that here we are substituting in \eqref{ThetaSoln}. Recall the signs here correspond to $\det({\sf O})=\pm 1$. Equation \eqref{VarphiDefn} is transcedental. We plot $F^{+}_{r}(\varphi)$ in Fig.~\ref{FactorizationA} as a function of $r$ and $\varphi$. 
%
%
We find that for $\det({\sf O})=1$
\begin{align}
	{}& \Im\left[ \frac{\rt{\theta} \cos \varphi + i \rt{1-\theta} \sin \varphi}
	               {\rt{\theta} \sin \varphi - i \rt{1-\theta} \cos \varphi} \right]  \nn \\
	               {}& = \frac{\rt{\theta(1-\theta)}}{\theta \sin^2 \varphi + (1-\theta) \cos^2 \varphi} \; \ge 0 \;,
\end{align}
which means the range of $F^{+}_r$ should be between $0$ and $\pi$ as seen in Fig.~\ref{FactorizationA}. A noteworthy feature occurs at $r=1$. From \eqref{ThetaSoln} we get $\theta_{1}=1/2$ which gives
\begin{align}
\label{F1}
	F^{+}_{1} = \arg\left[ \frac{\cos \varphi + i \sin \varphi}{\sin \varphi - i \cos \varphi} \right] = \arg[i] = \pi/2 \;.
\end{align}

\begin{figure*}
\includegraphics[width=17.5cm]{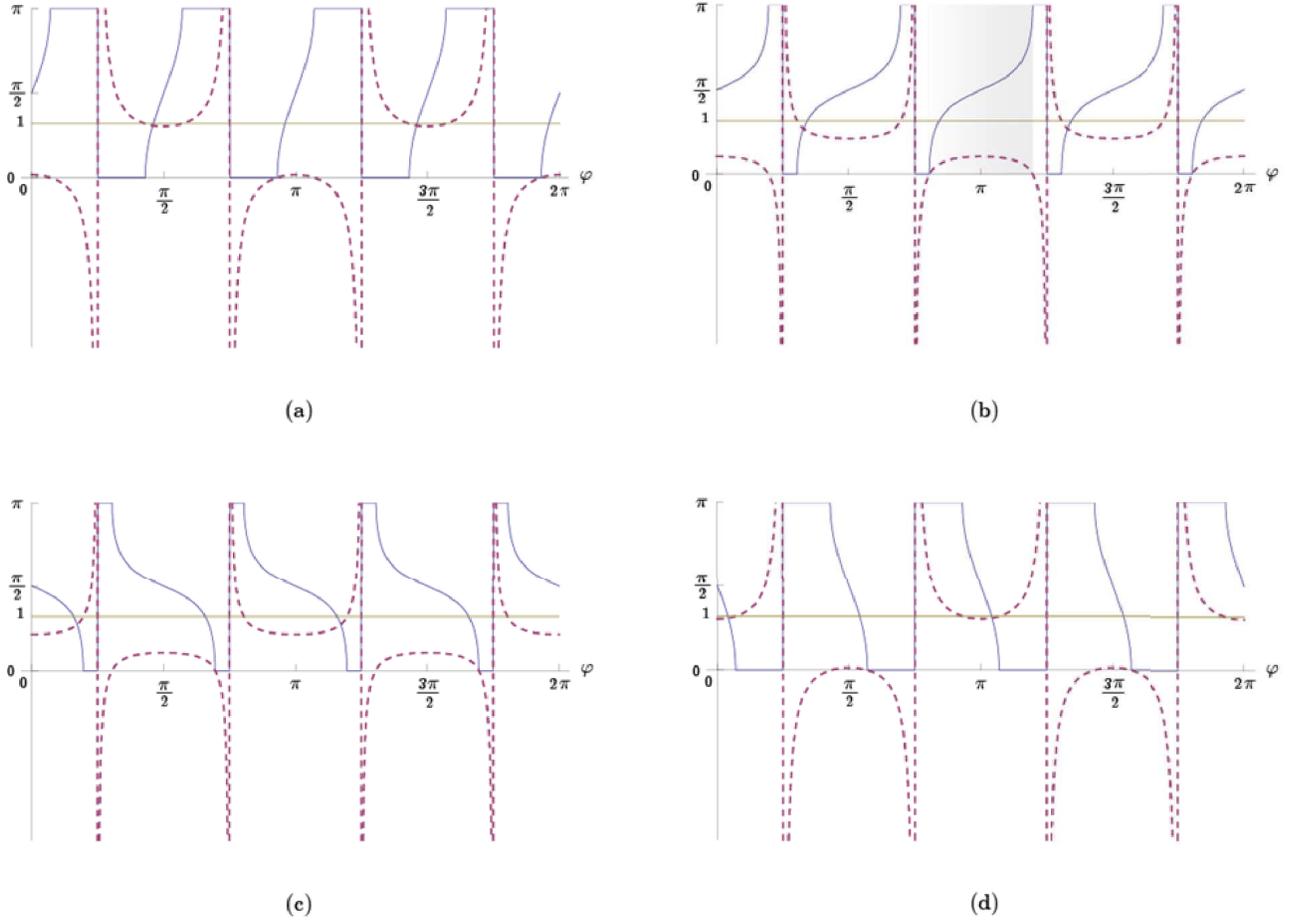} 
\caption{Plots of $F^{+}_{r}(\varphi)$ (solid blue line) and $\theta_r(\varphi)$ (dashed purple line) for selected values of $r$: (a) $r=0.05$, (b) $r=0.5$ (c) $r=2$, (d) $r=20$.}
\label{ThetaAndFPlots}
\end{figure*}

We also plot $F^{+}_{r}(\varphi)$ (solid blue line) together with $\theta_r(\varphi)$ (purple dashed line) for selected values of $r$ as a function of $\varphi$ in Fig.~\ref{ThetaAndFPlots}.
Observe that $F^{+}_r(\varphi)$ always has a support which corresponds to $\theta_r \in [0,1]$. We illustrate this region in Fig.~\ref{ThetaAndFPlots}(b) with a light shade. It can be seen in Fig.~\ref{ThetaAndFPlots} that this region persists for the various values of $r$ that we have chosen. This illustrates that there is always a value of $\varphi$ which solves \eqref{VarphiDefn} for $\delta \in [0,\pi]$, \emph{and} for which $\theta \in [0,1]$. Note that more than one value of $\varphi$ will solve \eqref{VarphiDefn} for a given $\delta$ which means that \mrep\ can be factorized by more than one ${\sf O}$ (recall \eqref{2X2O} and \eqref{MgivenBforL=1}). Having obtained a value of $\varphi$ we can then substitute it back into either \eqref{EqnForalpha1} or \eqref{EqnForalpha2} to obtain $\phi$.

In the above we have concentrated on the case $\delta \in [0,\pi]$. Here we show how \mrep\ can be factorized by \eqref{MOrthogonal} if we were given $\delta \in [-\pi,0]$. The analysis up to \eqref{VarphiDefn} would remain the same but in this case we can first let $\alpha'_2 \equiv \alpha_2 \pm \pi$ and solve 
\begin{align}
\label{DefnOfVarphiPrime}
	\alpha_1 - \alpha'_2 = F^{+}_{r}(\varphi')  
\end{align}
for $\varphi'$. The factorization of ${\sf M}\tp$ is then given by
\begin{align}
\label{MFactoredWithZ}
	\tbo{|m_1| e^{i\alpha_1}}{|m_2| e^{i\alpha_2}} 
	= {}& \tbo{|m_1| e^{i\alpha_1}}{|m_2| e^{i(\alpha'_2 \mp \pi)}}  \nn  \\
  = {}& \tbt{1}{0}{0}{-1} \tbo{|m_1| e^{i\alpha_1}}{|m_2| e^{i\alpha'_2}}  \;. 
\end{align}
where $\obt{|m_1| e^{i\alpha_1}}{|m_2| e^{i\alpha'_2}}\tp$ is factorized by
\begin{align}                                                 
  \tbo{|m_1| e^{i\alpha_1}}{|m_2| e^{i\alpha'_2}} 
  = \rt{\hbar\;\!\eta} \, e^{i\phi} \tbt{\cos \varphi'}{-\sin \varphi'}{\sin \varphi'}{\cos \varphi'} \tbo{\rt{\theta}}{-i\rt{\bar{\theta}}}  \;.
\end{align}
Substituting this into \eqref{MFactoredWithZ} and taking the transpose we see that ${\sf M}$ has the form of \eqref{MOrthogonal} with the orthogonal matrix ${\sf O}$ given by  
\begin{align}
\label{detOMinusOne}
	{\sf O} = \tbt{\cos \varphi'}{\sin \varphi'}{-\sin \varphi'}{\cos \varphi'}  \tbt{1}{0}{0}{-1}  \;,
\end{align}
where $\varphi'$ defined by \eqref{DefnOfVarphiPrime}. Note that \eqref{detOMinusOne} now has a determinant of minus one. We see then for $\delta \in [-\pi,0]$ \mrep\ is still factorized by \eqref{MOrthogonal} but with an ${\sf O}$ such that $\det({\sf O})=-1$.

%
%

\end{document}